\documentclass[english]{revtex4}
\usepackage[T1]{fontenc}
\usepackage[latin1]{inputenc}
\usepackage{float}
\usepackage{graphicx}
\makeatletter

\usepackage{float}
\usepackage{graphicx}

\usepackage{babel}
\makeatother
\begin{document}
\begin{flushright}JLAB-THY-06-599 \\
\end{flushright}

\begin{flushright}\vspace{2cm}\end{flushright}


\begin{center}
 
{\Large \bf Weak Deeply Virtual Compton Scattering}

\vspace{4mm}

{\sc A. Psaker}$^{1,2}$, {\sc W. Melnitchouk}$^2$,  and 
{\sc A.V. Radyushkin}$^{1,2\dagger}$

\footnotetext[0]{$^{\dagger}$Also  at Bogoliubov Laboratory of Theoretical Physics, JINR, Dubna, Russian Federation} 

\vspace{2mm}

$^1${\it Physics Department, Old Dominion University,\\
		Norfolk, VA 23529} \\
\vspace{2mm}
$^2${\it Jefferson Lab, 12000 Jefferson Avenue,\\
		Newport News, VA 23606}
 
\end{center}

\begin{abstract}
\vspace{2cm}

We extend the analysis of   the 
deeply virtual Compton scattering process to the weak
interaction sector in the generalized Bjorken limit.  The virtual Compton
scattering amplitudes for the weak neutral and charged currents are
calculated at  the leading twist within the framework of the nonlocal
light-cone expansion via coordinate space QCD string operators.  Using a
simple model, we estimate  cross sections for neutrino scattering off
the nucleon, relevant for future high intensity neutrino beam facilities.

\vspace{5mm}

PACS number(s): 13.15.+g, 13.40.-f, 13.40.Gp, 13.60.-r, 13.60.Fz
\end{abstract}
\maketitle
\newpage

\section{Introduction}

As hybrids of form factors, parton distribution functions and
distribution amplitudes,  the generalized parton distributions (GPDs)
\cite{Muller:1998fv,Ji:1996ek,Radyushkin:1996nd,Radyushkin:1996ru,Ji:1996nm,Radyushkin:1997ki} provide the most complete
and unified description of hadronic structure
(for recent   reviews,   see \cite{Goeke:2001tz,Diehl:2003ny,Belitsky:2005qn} ).
Parton distributions
parameterize the longitudinal momentum distributions
(in an infinite momentum frame) of partons in the nucleon, while
the Fourier transforms of form factors in impact parameter space
describe the transverse coordinate distributions of the nucleon's
constituents \cite{Burkardt:2000za,Burkardt:2002hr}.
GPDs, on the other hand, simultaneously encapsulate both the
longitudinal momentum and transverse coordinate distributions,
and hence provide a much more comprehensive, three-dimensional
snapshot of the substructure of the nucleon.

At the same time, GPDs have received considerable attention in recent
years in connection with the so-called ``proton spin crisis''
\cite{Lampe:1998eu,Filippone:2001ux,Bass:2004xa}.
Namely, certain low moments of GPDs can be related to the total
angular momentum carried by quarks and gluons (or generically, partons)
in the nucleon \cite{Ji:1996ek}.
Combined with measurements of the quark helicity from inclusive deep
inelastic scattering, knowledge of GPDs can thus unravel the orbital
angular momentum carried by partons, on which little or no information
is currently available.

Typically, GPDs can be measured in hard exclusive lepton-production
processes, such as deeply virtual Compton scattering (DVCS),
\begin{eqnarray}
e^{-}\left(k\right)N\left(p_{1}\right) & \longrightarrow &
e^{-}\left(k'\right)N
\left(p_{2}\right)\gamma\left(q_{2}\right)\ .
\label{eq:DVCSreaction}
\end{eqnarray}
Here, an electron (or muon) scatters off a nucleon via the exchange
(in leading order QED) of a space-like photon with virtuality
$q_{1}^{2}=\left(k-k'\right)^{2}<0$, producing an intact nucleon
(with altered momentum) and a real photon in the final state.
(The four-momenta of the particles are denoted in the parentheses.)
At the quark level, in the leading-twist approximation,
the electromagnetic current couples to different quark species with
strength proportional to the squares of the quark charges, selecting
specific linear combinations of GPDs. Flavor-specific GPDs can be
reconstructed by considering DVCS from different hadrons (protons and
neutrons, for instance), and using isospin symmetry to
relate GPDs in the proton to those in the neutron.

On the other hand, different combinations of quark flavors can be
accessed by utilizing the weak current, which couples to quarks with
strengths proportional to the quark weak charges.
In analogy with DVCS, these can be studied in neutrino-induced
virtual Compton scattering,
\begin{eqnarray}
\nu\left(k\right)N\left(p_{1}\right) & \longrightarrow &
e^{-}\left(k'\right)N'
\left(p_{2}\right)\gamma\left(q_{2}\right)
\label{eq:CCreaction}
\end{eqnarray}
for the charged current, and
\begin{eqnarray}
\nu\left(k\right)N\left(p_{1}\right) & \longrightarrow &
\nu\left(k'\right)N
\left(p_{2}\right)\gamma\left(q_{2}\right)
\label{eq:NCreaction}
\end{eqnarray}
for neutral current reactions (and similarly for antineutrinos).
In particular, neutrino-induced DVCS can be more sensitive to the
$d$ quark content of the proton, in contrast to electromagnetic
probes which, because of the quark charges, are sensitive mostly
to the $u$ quark.

Because of the $V-A$ nature of the weak interactions, one can
probe $C$-odd combinations of GPDs as well as $C$-even (where $C$
is the charge conjugation operator), and thus measure independently
both the valence and sea content of GPDs. This has a particularly
novel application in the case of the polarized distributions.
The usual way to obtain information on the spin structure of the
nucleon is through inclusive deep inelastic scattering of a
polarized lepton from a polarized target. To separate the valence
and sea spin contributions one could imagine measuring the $C$-odd
polarized structure function by scattering neutrinos from a polarized
target, which would be rather prohibitive using existing technology
given the large volume of target needed to polarize. Because it is
sensitive to both spin-averaged and spin-dependent GPDs,
neutrino-induced DVCS could allow one to extract the spin-dependent
valence and sea quark distributions using {\em only} an unpolarized
target.

The weak current also allows one to study flavor nondiagonal GPDs,
such as those associated with the neutron-to-proton transitions in
charged current reactions in Eq.~(\ref{eq:CCreaction}).
The use of weak currents can thus provide an important tool to
complement the study of GPDs in the more familiar electron-induced
DVCS or deeply exclusive meson production processes.

Recently, neutrino scattering off nucleons for neutral currents was
discussed in Ref.~\cite{Amore:2004ng}, where the authors presented
the leading twist behavior of the cross section for the deeply virtual
neutrino scattering process.
In Ref.~\cite{Coriano:2004bk}, the analysis was extended to charged
currents, and the leading twist amplitude computed.
The  neutrino-induced hard exclusive production of  $D_s$ mesons
was  considered within   the GPD formalism in   
Ref.~\cite{Lehmann-Dronke:2001wu}.

In this paper, we present a comprehensive analysis of the weak
deeply virtual Compton scattering (wDVCS) processes in
Eqs.~(\ref{eq:CCreaction}) and (\ref{eq:NCreaction}), and give
a detailed account of the charged and neutral current amplitudes
and cross sections in the kinematics relevant to future high-intensity
neutrino experiments \cite{Drakoulakos:2004gn}.
Some of the formal results in this paper have appeared in
a preliminary report in Ref.~\cite{Psaker:2004sf}.
In Section~\ref{amplitudes}, we provide a detailed derivation of the 
leading twist weak neutral and charged current amplitudes using the 
nonlocal light-cone operator product expansion. We introduce an 
appropriate set of GPDs, which parameterize the wDVCS reactions. 
Together with the standard electromagnetic DVCS process, the latter
are analyzed in Section~\ref{processes}, where we also discuss the
relevant cross sections and kinematics. Here we closely follow the analysis 
of Ref.~\cite{Belitsky:2001ns}, however, we only keep contributions 
up to the twist-2 accuracy. Using a simple model for nucleon GPDs, which
includes only the valence quark contribution, we estimate the cross
sections for weak DVCS processes, and further compare the respective
rates in neutrino scattering with those in the standard electromagnetic
DVCS process.
Finally, we draw some conclusions and discuss future prospects in
Section~\ref{conclusions}.

\section{Weak virtual Compton scattering amplitude\label{amplitudes}}

In this section, we present a detailed derivation of the amplitudes
for weak virtual Compton scattering.
We begin with an analysis of some general aspects of the amplitudes,
before turning to the specific cases of the weak neutral and charged
currents.

\subsection{Generalities}

In analogy with the photon-induced DVCS amplitude, the weak virtual
Compton scattering amplitude can be obtained by replacing the incoming
virtual photon with the weak boson \emph{B},
\begin{eqnarray}
B\left(q_{1}\right)N\left(p_{1}\right) & \longrightarrow & \gamma
\left(q_{2}\right)N'\left(p_{2}\right)\ ,
\label{eq:weakDVCSsubprocess}
\end{eqnarray}
where $B=Z^{0}\;\mathrm{or}\; W^{\pm}$. In the case of the charged
$W^\pm$ bosons, the initial and final nucleons will be different.
In the Bjorken regime, where the virtuality of the initial boson
and the total center of mass energy squared of the virtual weak
boson--nucleon system are sufficiently large, namely $-q_{1}^{2}$ and
$\left(p_{1}+q_{1}\right)^{2} \rightarrow\infty$ with the ratio
$x_{B}\equiv-q_{1}^{2}/\left[2\left(p_{1}\cdot q_{1}\right)\right]$
finite, the relevant amplitude is dominated by light-like distances.
The dominant light-cone singularities, which generate the leading
power contributions in $1/\left|q_{1}^{2}\right|$ to the amplitude,
are represented by the so-called ``handbag diagrams'' in Fig.~1.
At leading twist and to the lowest order in $\alpha_{s}$, there are
two diagrams which contribute, in which the (hard) quark propagator
is convoluted with the (soft) four-point function parameterized in
terms of GPDs.
In addition, keeping the momentum transfer squared to the nucleon,
$t\equiv\left(p_{1}-p_{2}\right)^{2}$, as small as possible,
one arrives at the relevant kinematics required to study DVCS.
One of the methods to study the amplitude in these kinematics is
based on the nonlocal light-cone expansion of the product of currents
in QCD ``string'' operators in coordinate space
 \cite{Balitsky:1987bk},  which 
we will employ in the present work.
\begin{figure}[H]
\begin{center}
\includegraphics[%
  scale=0.7]{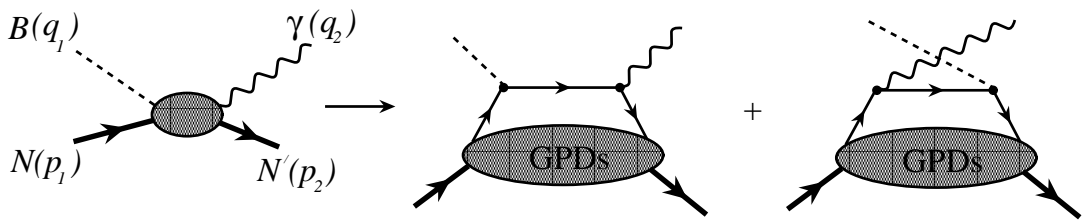}
\end{center}\caption{Weak deeply virtual Compton scattering of a
	weak vector boson $B$ from a nucleon $N$, producing a
	real photon $\gamma$ and recoil baryon $N'$ in the final
	state.  In the leading ``handbag'' approximation (two
	rightmost diagrams), the boson scatters from a single
	quark in the nucleon, which emits the real photon.}
\label{weakhandbag1}
\end{figure}

In the most general nonforward case, the virtual Compton scattering
amplitude is given by a Fourier transform of the correlation function
of two electroweak currents. In particular, for the standard virtual
Compton process on the nucleon, both currents ($J_{EM}^\mu$ and
$J_{EM}^\nu$) are electromagnetic and the amplitude can be written as:
\begin{eqnarray}
T_{EM}^{\mu\nu} & = & i
\int d^{4}x\int d^{4}y\; e^{-iq_{1}\cdot x+iq_{2}\cdot y}\left
\langle N\left(p_{2},s_{2}\right)\right|T\left\{ J_{EM}^{\mu}
\left(y\right)J_{EM}^{\nu}\left(x\right)\right\} \left|N
\left(p_{1},s_{1}\right)\right\rangle\ .
\label{eq:standardamplitude}
\end{eqnarray}
Here $p_1$ and $p_2$ are the four-momenta of the initial and final
nucleons, and $s_1$ and $s_2$ are their spins.
Similarly, for the weak process, with an incoming $Z^0$ or $W^\pm$
boson and outgoing photon, the amplitude is:
\begin{eqnarray}
T_{W}^{\mu\nu} & = & i\int d^{4}x
\int d^{4}y\; e^{-iq_{1}\cdot x+iq_{2}\cdot y}\left
\langle N'\left(p_{2},s_{2}\right)\right|T\left\{ J_{EM}^{\mu}
\left(y\right)J_{W}^{\nu}
\left(x\right)\right\} \left|N\left(p_{1},s_{1}
\right)\right\rangle\ .
\label{eq:weakComptonamplitude1}
\end{eqnarray}
Here $J_{W}^{\nu}\left(x\right)$ corresponds to either the weak
neutral current $J_{WN}^{\nu}\left(x\right)$, or the weak charged
current $J_{WC}^{\nu}\left(x\right)$. Note that in the electromagnetic
and weak neutral cases both the incoming and outgoing nucleons are
the same, $N=N'$, whereas for the charged current case they are
different, $N \neq N'$.

It will be convenient in the analysis to use symmetric coordinates,
defined by introducing center and relative coordinates of the points
\emph{x} and \emph{y}, $X\equiv\left(x+y\right)/2$ and $z\equiv y-x$.
The weak virtual Compton scattering amplitude then takes the form:
\begin{eqnarray}
T_{W}^{\mu\nu} & = & i\int d^{4}X\int d^{4}z\;
e^{-i\left(q_{1}-q_{2}\right)\cdot X+i
\left(q_{1}+q_{2}\right)\cdot z/2}\left
\langle N'\left(p_{2},s_{2}\right)
 \right|T\left\{
J_{EM}^{\mu}\left(X+z/2\right)J_{W}^{\nu}\left(X-z/2\right)\right\}
\left|N\left(p_{1},s_{1}\right)\right\rangle.
\label{eq:weakComptonamplitude2}
\end{eqnarray}
Furthermore, in order to treat the initial and final hadrons in a
symmetric manner, we introduce as independent momentum variables the
averages of the boson and hadron momenta,
$q\equiv\left(q_{1}+q_{2}\right)/2$ and
$p\equiv\left(p_{1}+p_{2}\right)/2$, and the overall momentum
transfer, $r\equiv p_{1}-p_{2}=q_{2}-q_{1}$. Accordingly, we have:
\begin{eqnarray}
q^{2}=q_{1}^{2}/2-t/4 & \mathrm{with} & t=r^{2}.
\label{eq:qsquared}
\end{eqnarray}
From the on-mass-shell conditions, $p_{1}^{2}=M_{1}^{2}$ and
$p_{2}^{2}=M_{2}^{2}$, one then has:
\begin{eqnarray}
p^{2}=\frac{1}{2}\left(M_{1}^{2}+M_{2}^{2}-t/2
\right) & \mathrm{and} &
p\cdot r=\frac{1}{2}\left(M_{1}^{2}-M_{2}^{2}\right)\ ,
\label{eq:frommassshellconditions1}
\end{eqnarray}
where $M_{1}$ and $M_{2}$ denote the masses of the initial and
final nucleons, respectively. In the following we shall neglect
the mass difference between the proton and neutron, and set
$M_{1}=M_{2}\equiv M (\simeq0.94\;\mathrm{GeV})$, in which case
these relations simplify to:
\begin{eqnarray}
p^{2}=M^{2}-t/4 & \mathrm{and} & p\cdot r = 0\ .
\label{eq:frommassshellconditions2}
\end{eqnarray}
After translating to the center coordinates in
Eq.~(\ref{eq:weakComptonamplitude2}),
$\left\langle p_{2}\right|J^{\mu}\left(X\right)\left|p_{1}
\right\rangle=\left\langle p_{2}\right|J^{\mu}\left(0\right)
\left|p_{1}\right\rangle e^{-i\left(p_{1}-p_{2}\right)\cdot X}$,
and integrating over $X$, one can write the weak amplitude as:
\begin{eqnarray}
T_{W}^{\mu\nu} & = & \left(2\pi\right)^{4}\delta^{\left(4\right)}
\left(p_{1}+q_{1}-p_{2}-q_{2}
\right)\mathsf{\mathcal{T}}_{W}^{\mu\nu}\ ,
\label{eq:weakComptonamplitude3}
\end{eqnarray}
where the {\em reduced} weak virtual Compton scattering amplitude is:
\begin{eqnarray}
\mathsf{\mathcal{T}}_{W}^{\mu\nu} & = & i\int d^{4}z\; e^{iq\cdot z}
\left\langle N'\left(p-r/2,s_{2}\right)\right|T\left\{J_{EM}^{\mu}
\left(z/2\right)J_{W}^{\nu}\left(-z/2\right)\right\} \left|N
\left(p+r/2,s_{1}\right)\right\rangle\ .
\label{eq:reducedamplitude}
\end{eqnarray}
The latter appears in the invariant matrix element (i.e. in the
T-matrix) of the specific wDVCS process, and will be computed at
the twist-2 level in the DVCS kinematics defined above.
This approximation amounts to neglecting contributions of the order
$M^{2}/q^{2}$ and $t/q^{2}$.
Since the final state photon is on-shell, $q_{2}^{2}=0$, it follows
in this particular kinematics that
$r\cdot q_{1} \simeq -q_{1}^{2}/2 = x_{B}\left(p_{1} \cdot q_{1}\right)$.
Hence the momentum transfer should have a large component in the
direction of the average nucleon momentum,
\begin{eqnarray}
r & = & 2\eta p+\Delta\ ,
\label{eq:momentumtransfer}
\end{eqnarray}
characterized by the skewness parameter
\begin{eqnarray}
\eta & \equiv & \frac{r\cdot q}{2\left(p\cdot q\right)}\ .
\label{eq:etaparameter}
\end{eqnarray}
The remainder $\Delta$ in Eq.~(\ref{eq:momentumtransfer}) is
transverse to both $p$ and $q$
\cite{Radyushkin:2000jy,Radyushkin:2000ap}. Moreover, in the
DVCS kinematics, $\eta$ coincides with the scaling variable
$\xi$ \cite{Ji:1996ek}, where
\begin{eqnarray}
\xi & \equiv & -\frac{q^{2}}{2\left(p\cdot q\right)}\ .
\label{eq:ksiparameter}
\end{eqnarray}
It is then easy to verify that in the limit $t/q^{2} \to 0$,
one has $\xi=x_{B}/\left(2-x_{B}\right)$.

\begin{figure}[H]
\begin{center}
\includegraphics[%
  scale=0.6]{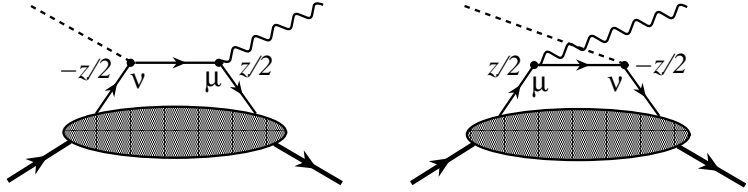}
\end{center}\caption{Handbag diagrams for the weak deeply virtual
	Compton scattering amplitude in the $s$-channel
	(left diagram) and $u$-channel (right diagram).}
\label{weakhandbag2}
\end{figure}

Having introduced the reduced wDVCS amplitude
$\mathsf{\mathcal{T}}_{W}^{\mu\nu}$ and the DVCS kinematics, we now
turn to the formal light-cone expansion of the time-ordered product
$T\left\{ J_{EM}^{\mu}\left(z/2\right)J_{W}^{\nu}
\left(-z/2\right)\right\}$ in the coordinate representation.
The expansion is performed in terms of QCD string operators \cite{Balitsky:1987bk}. The string operators
have gauge links along the straight line between the fields, however,
for brevity we will not write them explicitly. The leading light-cone
singularity is given by the $s$- and $u$-channel handbag diagrams
shown in Fig.~\ref{weakhandbag2}. The hard part of each of the
diagrams begins at zeroth order in $\alpha_{s}$ with the purely
tree level diagrams, in which the virtual weak boson and real photon
interact with the (massless) quarks. The free quark propagator
$S(z)$ between the initial and final quark fields in the coordinate
representation is given by:
\begin{eqnarray}
\not\! S\left(z\right)=\frac{\not\! z}{2\pi^{2}
\left(z^{2}-i0\right)^{2}} & = & \int
\frac{d^{4}l}{\left(2\pi\right)^{4}}\; e^{-il\cdot z}
\frac{\not l}{l^{2}+i0}\ .
\label{eq:freequarkpropagator}
\end{eqnarray}
Since the weak current couples to the quark fields through two types
of vertices, $qqZ^{0}$ and $qqW^{\pm}$, the quark fields at
coordinates $\pm z/2$ can carry either the same or different flavor
quantum numbers. We shall treat these two cases separately.
%

\subsection{Weak neutral amplitude}

In this section we expand the time-ordered product of the weak neutral
and electromagnetic currents. Omitting the overall vertex factor
$-\left|e\right|g/\cos\theta_{W}$, where $g$ is the weak coupling
constant, and $\theta_{W}$ is the Weinberg angle, one has:
\begin{eqnarray}
iT\left\{ J_{EM}^{\mu}\left(z/2\right)J_{WN}^{\nu}\left(-z/2\right)
\right\}  & = & i\sum_{f}Q_{f}\left[\bar{\psi}_{f}\left(z/2\right)
\gamma^{\mu}i\not\! S\left(z\right)\gamma^{\nu}\frac{1}{2}
\left(c_{V}^{f}-\gamma_{5}c_{A}^{f}\right)\psi_{f}
\left(-z/2\right)\right.\nonumber \\
 &  & \left.+\bar{\psi}_{f}\left(-z/2\right)
\gamma^{\nu}\frac{1}{2}\left(c_{V}^{f}-
\gamma_{5}c_{A}^{f}\right)i\not\! S\left(-z\right)
\gamma^{\mu}\psi_{f}\left(z/2\right)\right]\ ,
\label{eq:weakneutralexpansion1}
\end{eqnarray}
where $Q_{f}$ denotes the electric charge of the quark with flavor
$f$ (in units of $\left|e\right|$). In the Standard Model,
the weak vector and axial vector charges are given by:
\begin{eqnarray}
c_{V}^{u,c,t}=1/2-2Q_{u,c,t}\sin^{2}\theta_{W} & \mathrm{and}
& c_{V}^{d,s,b}=-1/2-2Q_{d,s,b}
\sin^{2}\theta_{W}\ , \\
c_{A}^{u,c,t}=1/2 & \mathrm{and} & c_{A}^{d,s,b}=-1/2\ ,
\label{eq:weakneutralcharges}
\end{eqnarray}
with $\sin^{2}\theta_{W}\simeq0.23$. Using the $\gamma$-matrix
formula: 
\begin{eqnarray}
\gamma^{\mu}\gamma^{\rho}\gamma^{\nu} & = &
\left(s^{\mu\rho\nu\eta}+i\epsilon^{\mu\rho\nu\eta}
\gamma_{5}\right)\gamma_{\eta}\ ,
\label{eq:gammamatrixformula}
\end{eqnarray}
where $s^{\mu\rho\nu\eta}\equiv g^{\mu\rho}g^{\nu\eta}+
g^{\mu\eta}g^{\rho\nu}-g^{\mu\nu}g^{\rho\eta}$ is the symmetric
tensor and $\epsilon^{\mu\rho\nu\eta}$ the antisymmetric 
tensor \footnote{We adopt the convention $\epsilon_{0123}=1$.}  in
Lorentz indices ($\mu,\nu$), the original bilocal quark operators
can be expressed in terms of the vector and axial vector string
operators with only one uncontracted Lorentz index:
\begin{eqnarray}
\label{eq:uncontractedstringoperators1}
\mathcal{O}_{\eta}^{f\pm}\left(z\left|0\right.\right) & \equiv &
\left[\bar{\psi}_{f}\left(z/2\right)\gamma_{\eta}\psi_{f}
\left(-z/2\right)\pm\left(z\rightarrow-z\right)\right]\ , \\
\mathcal{O}_{5\eta}^{f\pm}\left(z\left|0\right.\right) & \equiv &
\left[\bar{\psi}_{f}\left(z/2\right)\gamma_{\eta}
\gamma_{5}\psi_{f}\left(-z/2\right)\pm
\left(z\rightarrow-z\right)\right]\ .
\label{eq:uncontractedstringoperators2}
\end{eqnarray}
Accordingly, the time-ordered product of currents in
Eq.~(\ref{eq:weakneutralexpansion1}) assumes the form:
\begin{eqnarray}
iT\left\{ J_{EM}^{\mu}\left(z/2\right)J_{WN}^{\nu}
\left(-z/2\right)\right\}
& = & -\frac{z_{\rho}}{4\pi^{2}z^{4}}
\sum_{f}Q_{f}\left\{ c_{V}^{f}\left[s^{\mu\rho\nu\eta}
\mathcal{O}_{\eta}^{f-}\left(z\left|0\right.\right)-i
\epsilon^{\mu\rho\nu\eta}\mathcal{O}_{5\eta}^{f+}
\left(z\left|0\right.\right)\right]\right.\nonumber \\
&  & \left.-\ c_{A}^{f}\left[s^{\mu\rho\nu\eta}\mathcal{O}_{5\eta}^{f-}
\left(z\left|0\right.\right)-i\epsilon^{\mu\rho\nu\eta}
\mathcal{O}_{\eta}^{f+}\left(z\left|0\right.\right)\right]\right\}\ .
\label{eq:weakneutralexpansion2}
\end{eqnarray}
In contrast to the standard electromagnetic DVCS process, here we have
two additional terms. Namely, the presence of the axial part
$\gamma_{5}c_{A}^{f}$ of the \emph{V--A} interaction gives rise to a
vector current symmetric in ($\mu, \nu$), and to an axial vector
current antisymmetric in ($\mu, \nu$).

The string operators in Eqs.~(\ref{eq:uncontractedstringoperators1})
and (\ref{eq:uncontractedstringoperators2}) do not have a definite twist.
To isolate their twist-2 parts, one uses a Taylor series expansion in the
relative coordinate $z$.
The expansion gives rise to the local operators
$\bar{\psi}_{f}\left(0\right) \gamma_{\eta}D_{\mu_{1}}
 \cdots D_{\mu_{n}}\psi_{f}\left(0\right)$
and
$\bar{\psi}_{f}\left(0\right)\gamma_{\eta}\gamma_{5}D_{\mu_{1}}
 \cdots D_{\mu_{n}}\psi_{f}\left(0\right)$,
where $D_{\mu}$ is the covariant derivative.
To obtain the twist-2 contributions, one needs to project the totally
symmetric, traceless parts of the coefficients in the expansion,
which is effected by the following operation:
\begin{eqnarray}
\label{eq:symmetricparts1}
\left[\mathcal{O}_{\eta}^{f\pm}\left(z
\left|0\right.\right)\right]_{sym}
& = & \frac{\partial}{\partial z^{\eta}}\int_{0}^{1}d\beta\;
\left[\bar{\psi}_{f}\left(\beta z/2\right)\not\! z\psi_{f}
\left(-\beta z/2\right)\mp\left(z\rightarrow-z\right)
\right]\ , \\
\left[\mathcal{O}_{5\eta}^{f\pm}
\left(z\left|0\right.\right)\right]_{sym}
& = & \frac{\partial}{\partial z^{\eta}}\int_{0}^{1}d\beta\;
\left[\bar{\psi}_{f}\left(\beta z/2\right)\not\! z\gamma_{5}
\psi_{f}\left(-\beta z/2\right)\mp\left(z\rightarrow-z\right)\right]\ .
\label{eq:symmetricparts2}
\end{eqnarray}
The subtraction of traces is implemented by imposing the harmonic
condition on the string operators on the right-hand-sides of
Eqs.~(\ref{eq:symmetricparts1}) and (\ref{eq:symmetricparts2}).
In other words, the contracted operators
\begin{eqnarray}
\mathcal{O}^{f\pm}\left(z\left|0\right.\right) & \equiv &
\left[\bar{\psi}_{f}
\left(z/2\right)\not\! z\psi_{f}
\left(-z/2\right)\pm\left(z\rightarrow-z\right)\right]\ , \\
\mathcal{O}_{5}^{f\pm}\left(z\left|0\right.\right) & \equiv &
\left[\bar{\psi}_{f}
\left(z/2\right)\not\! z\gamma_{5}
\psi_{f}\left(-z/2\right)\pm\left(z\rightarrow-z\right)\right]
\label{eq:contractedoperatorsWN}
\end{eqnarray}
should satisfy the d'Alembert equation with respect to $z$:
\begin{eqnarray}
\partial^{2}_{z}\left[\mathcal{O}^{f\pm}\left(z\left|0\right.
\right)\right]_{\rm twist-2} & = & 0\ ,
\label{eq:d'Alembert}
\end{eqnarray}
and similarly for the twist-2 part of
$\mathcal{O}_{5}^{f\pm}\left(z\left|0\right.\right)$.

To compute the amplitude $\mathsf{\mathcal{T}}_{W}^{\mu\nu}$ in
Eq.~(\ref{eq:reducedamplitude}), we sandwich the contracted twist-2
operators between the initial and final nucleon states.
To construct a parametrization for the nonforward nucleon matrix
elements, we use a spectral representation, with the relevant spectral
functions corresponding to off-forward parton distributions
(OFPDs) \footnote{Here, and in the following, we will use this
	terminology rather than the more generic ``generalized
	parton distributions'' (GPDs), of which OFPDs are a specific
	example.}.
Since the coordinate $z$ runs over the whole four-dimensional space,
in principle the parametrization should be valid everywhere in $z$.
However, the inclusion of the $z^{2}$ terms in the matrix elements
generates $M^{2}/q^{2}$ and $t/q^{2}$ corrections to the amplitude
(analogous to the well-known target mass corrections in deep inelastic
scattering \cite{Nachtmann:1973mr,Georgi:1976ve,Steffens2006}), and
will hence be neglected.
It is sufficient therefore to provide a parametrization only on the
light-cone \cite{Radyushkin:2000ap}, namely:
\begin{eqnarray}
\label{eq:wnparametrization1}
\left\langle N\left(p_{2},s_{2}\right)\right|\mathcal{O}^{f\pm}
\left(z\left|0\right.\right)
\left|N\left(p_{1},s_{1}\right)\right\rangle_{z^2=0}
& = & \bar{u}\left(p_{2},s_{2}
\right)\not\! zu\left(p_{1},s_{1}\right)
\int_{-1}^{1}dx\; e^{ix p \cdot z}H_{f}^{\pm}
\left(x,\xi,t\right) \nonumber \\
&  & +\bar{u}\left(p_{2},s_{2}\right)
\frac{\left(\not\! z\not\! r-\not\! r\not\! z\right)}{4M}u
\left(p_{1},s_{1}\right)
\int_{-1}^{1}dx\; e^{ix p \cdot z}E_{f}^{\pm}
\left(x,\xi,t\right)\ , \\
\left\langle N\left(p_{2},s_{2}\right)\right|\mathcal{O}_{5}^{f\pm}
\left(z\left|0\right.\right)\left|N\left(p_{1},s_{1}\right)\right
\rangle_{z^2=0}  & = & \bar{u}\left(p_{2},s_{2}\right)\not\! z\gamma_{5}u
\left(p_{1},s_{1}\right)\int_{-1}^{1}dx\; e^{ix p \cdot z}
\widetilde{H}_{f}^{\mp}\left(x,\xi,t\right)\nonumber \\
&  & -\bar{u}\left(p_{2},s_{2}\right)\frac{\left(r\cdot z\right)}{2M}
\gamma_{5}u\left(p_{1},s_{1}\right)
\int_{-1}^{1}dx\; e^{ix p \cdot z}
\widetilde{E}_{f}^{\mp}\left(x,\xi,t\right)\ .
\label{eq:wnparametrization2}
\end{eqnarray}
The flavor dependent OFPDs in Eqs.~(\ref{eq:wnparametrization1}) and
(\ref{eq:wnparametrization2}) refer to the corresponding quark flavor
$f$ in the nucleon $N$.
They depend on the usual light-cone momentum fraction $x$, the skewness
parameter $\xi$, which specifies the longitudinal momentum asymmetry,
and the invariant momentum transfer $t$ to the target. As illustrated,
for example in the $s$-channel diagram of Fig.~\ref{weakhandbag2},
the parton taken out of the parent nucleon at the space-time point  ``-$z/2$''
carries a fraction $x+\xi$ of the average nucleon momentum $p$, while
the momentum of the reabsorbed parton at the space-time point ``$z/2$''  is
$\left(x-\xi\right)p$.

Note that
$\mathcal{O}_{5}^{f\pm}\left(z\left|0\right.\right)$
has a superscript opposite in sign with respect to the corresponding
``tilded'' 
distributions $\widetilde{H}_{f}^{\mp}\left(x,\xi,t\right)$ and
$\widetilde{E}_{f}^{\mp}\left(x,\xi,t\right)$. While the standard DVCS
process gives access only to the {\em plus} distributions (i.e. the sum
of quark and antiquark distributions), scattering via the virtual weak
boson exchange probes also the {\em minus} distributions. The latter
correspond to the difference in quark and antiquark distributions, or
the valence configuration. The advantage of using the plus and minus
distributions, as opposed to the usual OFPDs $H_{f}\left(x,\xi,t\right)$,
$E_{f}\left(x,\xi,t\right)$, $\widetilde{H}_{f}\left(x,\xi,t\right)$ and
$\widetilde{E}_{f}\left(x,\xi,t\right)$, which parametrize the matrix
elements
of operators $\bar{\psi}_{f}\left(z/2\right)\not\! z\psi_{f}
\left(-z/2\right)$ and $\bar{\psi}_{f}\left(z/2\right)\not\! z
\gamma_{5}\psi_{f}\left(-z/2\right)$, is that they both have well-defined
symmetry properties with respect to the scaling variable $x$. By
transforming $z\rightarrow-z$ and $x\rightarrow-x$, one can readily
establish the following crossing symmetry relations:
\begin{eqnarray}
\label{eq:symmetrypropertiesplusminusGPDs1}
H_{f}^{\pm}\left(x\right) & = & \mp H_{f}^{\pm}
\left(-x\right) , \\
E_{f}^{\pm}\left(x\right) & = & \mp E_{f}^{\pm}
\left(-x\right) , \\
\widetilde{H}_{f}^{\pm}\left(x\right) & = &
\pm\widetilde{H}_{f}^{\pm}\left(-x\right) , \\
\widetilde{E}_{f}^{\pm}\left(x\right) & = &
\pm\widetilde{E}_{f}^{\pm}\left(-x\right) .
\label{eq:symmetrypropertiesplusminusGPDs4}
\end{eqnarray}

Substituting the parametrizations (\ref{eq:wnparametrization1})
and (\ref{eq:wnparametrization2}) into the right-hand-sides of
Eqs.~(\ref{eq:symmetricparts1}) and (\ref{eq:symmetricparts2}),
one isolates the twist-2 terms by taking the derivative with respect
to $z$, and integrating by parts over the parameter $\beta$, keeping
only the surface terms with the arguments
$\bar{\psi}_{f}\left(\pm z/2\right)$ and $\psi_{f}\left(\pm z/2\right)$.
Finally, the integral over $z$ is carried out with the help of the
inversion formula for $\not\! S\left(z\right)$:
\begin{eqnarray}
\int d^{4}z\; e^{il\cdot z}
\frac{z_{\rho}}{2\pi^{2}\left(z^{2}-i0\right)^{2}} & = &
\frac{l_{\rho}}{\left(l^{2}+i0\right)}\ .
\label{eq:inversionformula}
\end{eqnarray}
Here the momentum $l$ is given by $l=\left(xp+q\right)$, so
that $l^{2}$ in the denominator of Eq.~(\ref{eq:inversionformula})
becomes $l^{2}=2\left(p\cdot q\right)\left(x-\xi\right)$.
The expression for the reduced weak neutral virtual Compton
scattering amplitude in the leading-twist approximation can then
be written (we will implicitly deal with twist-2 amplitudes
henceforth):
\begin{eqnarray}
\mathcal{T}_{WN}^{\mu\nu} & = & -\frac{1}{4\left(p\cdot q\right)}
\sum_{f}Q_{f}\int_{-1}^{1}
\frac{dx}{\left(x-\xi+i0\right)}\nonumber \\
&  & \times\Bigg\lbrace c_{V}^{f}\left[s^{\mu\rho\nu\eta}l_{\rho}
\left[\bar{u}\left(p_{2},s_{2}\right)\gamma_{\eta}u\left(p_{1},s_{1}
\right)H_{f}^{+}\left(x,\xi,t\right)+\bar{u}\left(p_{2},s_{2}
\right)\frac{\gamma_{\eta}\not\! r-\not\! r\gamma_{\eta}}{4M}u
\left(p_{1},s_{1}\right)E_{f}^{+}\left(x,\xi,t\right)
\right]\right.\nonumber \\
&  & \hspace*{0.5cm}\left.+i\epsilon^{\mu\nu\rho\eta}l_{\rho}
\left[\bar{u}\left(p_{2},s_{2}\right)\gamma_{\eta}
\gamma_{5}u\left(p_{1},s_{1}\right)\widetilde{H}_{f}^{+}
\left(x,\xi,t\right)-\bar{u}\left(p_{2},s_{2}\right)
\frac{r_{\eta}}{2M}\gamma_{5}u\left(p_{1},s_{1}\right)
\widetilde{E}_{f}^{+}\left(x,\xi,t\right)\right]
\right]\nonumber \\
&  & \hspace*{0.5cm}-c_{A}^{f}\left[s^{\mu\rho\nu\eta}l_{\rho}
\left[\bar{u}\left(p_{2},s_{2}\right)\gamma_{\eta}
\gamma_{5}u\left(p_{1},s_{1}\right)\widetilde{H}_{f}^{-}
\left(x,\xi,t\right)-\bar{u}\left(p_{2},s_{2}
\right)\frac{r_{\eta}}{2M}\gamma_{5}u\left(p_{1},s_{1}
\right)\widetilde{E}_{f}^{-}\left(x,\xi,t\right)
\right]\right.\nonumber \\
&  & \hspace*{0.5cm}\left.+i\epsilon^{\mu\nu\rho\eta}l_{\rho}
\left[\bar{u}\left(p_{2},s_{2}\right)
\gamma_{\eta}u\left(p_{1},s_{1}\right)H_{f}^{-}
\left(x,\xi,t\right)+\bar{u}\left(p_{2},s_{2}
\right)\frac{\gamma_{\eta}\not\! r-\not\! r\gamma_{\eta}}{4M}u
\left(p_{1},s_{1}\right)E_{f}^{-}\left(x,\xi,t\right)
\right]\right]\Bigg\rbrace\ .
\label{eq:weakneutralamplitude}
\end{eqnarray}
This can be further simplified by using the light-cone (Sudakov)
decomposition of the $\gamma$-matrix:
\begin{eqnarray}
\gamma^{\mu} & = & a^{\mu}\not\! n_{1}+b^{\mu}\not\! n_{2}+
\gamma_{\bot}^{\mu}\ ,
\label{eq:lightconedecomposition1}
\end{eqnarray}
where the four-vectors $n_{1}$ and $n_{2}$ in
Eq.~(\ref{eq:lightconedecomposition1}) are light-like,
$n_{1}^{2}=n_{2}^{2}=0$, and satisfy the condition
$n_{1}\cdot n_{2}=1$.
Identifying $n_{1}\leftrightarrow p$ and $n_{2}\leftrightarrow q_{2}$,
and neglecting the transverse component $\gamma_{\bot}^{\mu}$
(since it corresponds to the higher-twist contributions),
Eq.~(\ref{eq:lightconedecomposition1}) takes the form:
\begin{eqnarray}
\gamma^{\mu} & = & \frac{1}{\left(p\cdot q_{2}\right)}
\left(q_{2}^{\mu}\not\! p+p^{\mu}\not\! q_{2}\right)\ .
\label{eq:lightconedecomposition2}
\end{eqnarray}
Now, using the above decomposition, the Dirac equation,
$\not\!\! p_{1}u\left(p_{1},s_{1}\right)=\not\!\! p_{2}u
\left(p_{2},s_{2}\right)=0$
(recall that we neglect the nucleon mass),
and the symmetry properties in
Eqs.~(\ref{eq:symmetrypropertiesplusminusGPDs1})
-- (\ref{eq:symmetrypropertiesplusminusGPDs4}), the amplitude
(\ref{eq:weakneutralamplitude}) becomes:
\begin{eqnarray}
\mathcal{T}_{WN}^{\mu\nu} & = &
-\frac{1}{4\left(p\cdot q\right)}
\left\{ \left[
\frac{1}{\left(p\cdot q_{2}\right)}
\left(p^{\mu}q_{2}^{\nu}+p^{\nu}q_{2}^{\mu}\right)-g^{\mu\nu}
\right]\right.\nonumber \\
&  & \times\left[\mathcal{H}_{WN}^{+}\left(\xi,t\right)\bar{u}
\left(p_{2},s_{2}\right)
\not\! q_{2}u\left(p_{1},s_{1}\right)+
\mathcal{E}_{WN}^{+}\left(\xi,t\right)
\bar{u}\left(p_{2},s_{2}\right)
\frac{\left(\not\! q_{2}
\not\! r-\not\! r\not\! q_{2}\right)}{4M}u\left(p_{1},s_{1}\right)
\right.\nonumber \\
&  & \hspace*{0.5cm}
     \left.-\widetilde{\mathcal{H}}_{WN}^{-}\left(\xi,t\right)
\bar{u}\left(p_{2},s_{2}\right)\not\! q_{2}\gamma_{5}u
\left(p_{1},s_{1}\right)+
\widetilde{\mathcal{E}}_{WN}^{-}\left(\xi,t\right)
\frac{\left(q_{2}\cdot r\right)}{2M}
\bar{u}\left(p_{2},s_{2}\right)
\gamma_{5}u\left(p_{1},s_{1}\right)\right]\nonumber \\
&  & +\left[\frac{1}{\left(p\cdot q_{2}\right)}i
\epsilon^{\mu\nu\rho\eta}q_{2\rho}p_{\eta}
\right]\nonumber \\
&  & \times\left[\widetilde{\mathcal{H}}_{WN}^{+}\left(\xi,t\right)
\bar{u}\left(p_{2},s_{2}\right)\not\! q_{2}
\gamma_{5}u\left(p_{1},s_{1}\right)-\widetilde{\mathcal{E}}_{WN}^{+}
\left(\xi,t\right)
\frac{\left(q_{2}\cdot r\right)}{2M}
\bar{u}\left(p_{2},s_{2}\right)\gamma_{5}u\left(p_{1},s_{1}\right)
\right.\nonumber \\
&  & \hspace*{0.5cm}
     \left.\left.-\mathcal{H}_{WN}^{\mathit{-}}\left(\xi,t\right)
\bar{u}\left(p_{2},s_{2}
\right)\not\! q_{2}u\left(p_{1},s_{1}\right)-
\mathcal{E}_{WN}^{\mathit{-}}\left(\xi,t\right)
\bar{u}\left(p_{2},s_{2}\right)
\frac{\left(\not\! q_{2}\not\! r-\not\! r\not\! q_{2}
\right)}{4M}u\left(p_{1},s_{1}\right)
\right]\right\}\ ,
\label{eq:compactweakneutralamplitude}
\end{eqnarray}
where
\begin{eqnarray}
\mathcal{H}_{WN}^{+\left(-\right)}\left(\xi,t\right) &
\equiv & \sum_{f}Q_{f}c_{V\left(A\right)}^{f}\int_{-1}^{1}
\frac{dx}{\left(x-\xi+i0\right)}H_{f}^{+\left(-\right)}
\left(x,\xi,t\right) \nonumber \\
& = & \sum_{f}Q_{f}c_{V\left(A\right)}^{f}
\int_{-1}^{1}dx\; H_{f}\left(x,\xi,t\right)
\left(\frac{1}{x-\xi+i0}\pm\frac{1}{x+\xi-i0}\right)\;, \\
\mathcal{E}_{WN}^{+\left(-\right)}\left(\xi,t\right) &
\equiv & \sum_{f}Q_{f}c_{V\left(A\right)}^{f}\int_{-1}^{1}
\frac{dx}{\left(x-\xi+i0\right)}E_{f}^{+\left(-\right)}
\left(x,\xi,t\right) \nonumber \\
& = & \sum_{f}Q_{f}c_{V\left(A\right)}^{f}
\int_{-1}^{1}dx\; E_{f}\left(x,\xi,t\right)
\left(\frac{1}{x-\xi+i0}\pm\frac{1}{x+\xi-i0}
\right)\;, \\
\widetilde{\mathcal{H}}_{WN}^{+\left(-\right)}
\left(\xi,t\right) & \equiv &
\sum_{f}Q_{f}c_{V\left(A\right)}^{f}\int_{-1}^{1}
\frac{dx}{\left(x-\xi+i0\right)}\widetilde{H}_{f}^{+\left(-\right)}
\left(x,\xi,t\right) \nonumber \\
& = & \sum_{f}Q_{f}c_{V\left(A\right)}^{f}
\int_{-1}^{1}dx\;\widetilde{H}_{f}\left(x,\xi,t\right)
\left(\frac{1}{x-\xi+i0}\mp\frac{1}{x+\xi-i0}\right)\;, \\
\widetilde{\mathcal{E}}_{WN}^{+\left(-\right)}\left(\xi,t\right) &
\equiv & \sum_{f}Q_{f}c_{V\left(A\right)}^{f}\int_{-1}^{1}
\frac{dx}{\left(x-\xi+i0\right)}\widetilde{E}_{f}^{+\left(-\right)}
\left(x,\xi,t\right) \nonumber \\
& = & \sum_{f}Q_{f}c_{V\left(A\right)}^{f}
\int_{-1}^{1}dx\;\widetilde{E}_{f}\left(x,\xi,t\right)
\left(\frac{1}{x-\xi+i0}\mp\frac{1}{x+\xi-i0}\right)\;,
\label{eq:integralsofGPDsweakneutralcurrent}
\end{eqnarray}
is a new set of functions given by the integrals of OFPDs.
The amplitude $\mathcal{T}_{WN}^{\mu\nu}$ has both real and
imaginary parts, with the real part obtained using the principal
value prescription, and the imaginary part coming from the
singularity of the expression $1/\left(x\mp\xi\pm i0\right)$.
The latter generates the function $\delta\left(x\mp\xi\right)$,
which constrains evaluating OFPDs at the specific point $x=\pm\xi$.

\subsection{Weak charged amplitude}

For the weak charged amplitude, we write the expansion of the
time-ordered product of the weak charged and electromagnetic currents
(again omitting the overall vertex factor $-\left|e\right|g/\sqrt{2}$)
as:
\begin{eqnarray}
iT\left\{ J_{EM}^{\mu}\left(z/2\right)J_{WC}^{\nu}
\left(-z/2\right)\right\}  & = & -
\frac{z_{\rho}}{4\pi^{2}z^{4}}\sum_{f,f'}\left[Q_{f'}
\bar{\psi}_{f'}\left(z/2\right)
\gamma^{\mu}\gamma^{\rho}\gamma^{\nu}\left(1-\gamma_{5}\right)
\psi_{f}\left(-z/2\right)
\right.\nonumber \\
&  & \left.-Q_{f}\bar{\psi}_{f'}\left(-z/2\right)\gamma^{\nu}
\left(1-\gamma_{5}\right)\gamma^{\rho}\gamma^{\mu}\psi_{f}
\left(z/2\right)\right]\ .
\label{eq:weakchargedexpansion1}
\end{eqnarray}
Using the $\gamma$-matrix decomposition in
Eq.~(\ref{eq:gammamatrixformula}), and collecting terms,
one has: \footnote{Note that for the charged current one should in
	principle include the CKM matrix, since the quark mass
	eigenstates do not coincide with the weak eigenstates.
	In this work, however, we will set all Cabibbo angles to zero.}
\begin{eqnarray}
iT\left\{ J_{EM}^{\mu}\left(z/2\right)J_{WC}^{\nu}\left(-z/2
\right)\right\}
& = & -\frac{z_{\rho}}{4\pi^{2}z^{4}}\sum_{f,f'}
\Big\lbrace s^{\mu\rho\nu\eta}
\left[Q_{f'}\bar{\psi}_{f'}\left(z/2\right)\gamma_{\eta}
\psi_{f}\left(-z/2\right)-Q_{f}
\left(z\rightarrow-z\right)
\right]\nonumber \\
&  & -i\epsilon^{\mu\rho\nu\eta}
\left[Q_{f'}\bar{\psi}_{f'}\left(z/2
\right)\gamma_{\eta}\gamma_{5}\psi_{f}\left(-z/2\right)+
Q_{f}\left(z\rightarrow-z\right)
\right]\nonumber \\
&  & -s^{\mu\rho\nu\eta}\left[Q_{f'}\bar{\psi}_{f'}
\left(z/2\right)
\gamma_{\eta}\gamma_{5}\psi_{f}\left(-z/2\right)-Q_{f}
\left(z\rightarrow-z\right)
\right]\nonumber \\
&  & +i\epsilon^{\mu\rho\nu\eta}\left[Q_{f'}
\bar{\psi}_{f'}\left(z/2\right)
\gamma_{\eta}\psi_{f}\left(-z/2\right)+Q_{f}
\left(z\rightarrow-z\right)
\right]\Big\rbrace\ .
\label{eq:weakchargedexpansion2}
\end{eqnarray}
Here the sum over quark flavors is subject to an additional condition,
$Q_{f}-Q_{f'}=\pm1$, due to the fact that the virtual boson $W^{\pm}$
carries an electric charge, so that the initial and final nucleons
correspond to different particles.
In other words, the neutron-to-proton transition is via the exchange
of a $W^{+}$ using a neutrino beam, while the proton-to-neutron
transition is via $W^{-}$ using antineutrinos.
Moreover, the vector and axial vector string operators are not
diagonal in quark flavor, and are also accompanied by different
electric charges.
We can express the corresponding contracted string operators,
which appear when extracting the twist-2 part of the expansion
(\ref{eq:weakchargedexpansion2}), as the linear combinations:
\begin{eqnarray}
\label{eq:flavornoniagonalstringoperators1}
\left[Q_{f'}\bar{\psi}_{f'}\left(z/2\right)\not\! z\psi_{f}
\left(-z/2\right)\pm Q_{f}\left(z
\rightarrow-z\right)\right] & = & Q_{\pm}\mathcal{O}^{f'f+}
\left(z\left|0\right.\right)+Q_{\mp}\mathcal{O}^{f'f-}
\left(z\left|0\right.\right)\ , \\
\left[Q_{f'}\bar{\psi}_{f'}\left(z/2\right)\not\! z\gamma_{5}
\psi_{f}\left(-z/2\right)\pm Q_{f}
\left(z\rightarrow-z\right)\right] & = & Q_{\pm}
\mathcal{O}_{5}^{f'f+}
\left(z\left|0\right.\right)+Q_{\mp}\mathcal{O}_{5}^{f'f-}
\left(z\left|0\right.\right)\
\label{eq:flavornoniagonalstringoperators2}
\end{eqnarray}
of the operators:
\begin{eqnarray}
\label{eq:contractedoperatorsWC1}
\mathcal{O}^{f'f\pm}\left(z\left|0\right.\right) & \equiv &
\left[\bar{\psi}_{f'}
\left(z/2\right)\not\! z\psi_{f}\left(-z/2
\right)\pm\left(z\rightarrow-z\right)
\right]\ , \\
\mathcal{O}_{5}^{f'f\pm}\left(z\left|0\right.\right) &
\equiv & \left[\bar{\psi}_{f'}\left(z/2\right)
\not\! z\gamma_{5}\psi_{f}\left(-z/2\right)\pm
\left(z\rightarrow-z\right)\right]\ ,
\label{eq:contractedoperatorsWC2}
\end{eqnarray}
with the coefficients $Q_{\pm}=\left(Q_{f'}\pm Q_{f}\right)/2$.
The matrix elements of these operators are then parametrized in
terms of the flavor nondiagonal OFPDs:
\begin{eqnarray}
\label{eq:wcparametrization1}
\left\langle N'\left(p_{2},s_{2}
\right)\right|\mathcal{O}^{f'f\pm}
\left(z\left|0\right.\right)
\left|N\left(p_{1},s_{1}\right)
\right\rangle_{z^2=0}
& = & \bar{u}\left(p_{2},s_{2}\right)\not\! zu
\left(p_{1},s_{1}\right)
\int_{-1}^{1}dx\; e^{ix p \cdot z}H_{f'f}^{\pm}
\left(x,\xi,t\right)\nonumber \\
&  & +\bar{u}\left(p_{2},s_{2}\right)
\frac{\left(\not\! z\not\! r-\not\! r\not\! z\right)}{4M}u
\left(p_{1},s_{1}\right)
\int_{-1}^{1}dx\; e^{ix p \cdot z}E_{f'f}^{\pm}
\left(x,\xi,t\right)\ , \\
\left\langle N'\left(p_{2},s_{2}\right)
\right|\mathcal{O}_{5}^{f'f\pm}
\left(z\left|0\right.\right)\left|N\left(p_{1},s_{1}
\right)\right
\rangle_{z^2=0}  & = & \bar{u}\left(p_{2},s_{2}
\right)\not\! z\gamma_{5}u
\left(p_{1},s_{1}\right)
\int_{-1}^{1}dx\; e^{ix p \cdot z}
\widetilde{H}_{f'f}^{\mp}\left(x,\xi,t
\right)\nonumber \\
&  & -\bar{u}\left(p_{2},s_{2}\right)
\frac{\left(r\cdot z\right)}{2M}
\gamma_{5}u\left(p_{1},s_{1}\right)
\int_{-1}^{1}dx\; e^{ix p \cdot z}
\widetilde{E}_{f'f}^{\mp}\left(x,\xi,t\right)\ .
\label{eq:wcparametrization2}
\end{eqnarray}
These correspond to the situation in which a quark with flavor $f$
is removed from the target nucleon, and a quark with flavor $f'$ is
reabsorbed. Finally, with the help of
Eqs.~(\ref{eq:flavornoniagonalstringoperators1}) and
(\ref{eq:flavornoniagonalstringoperators2}) and the parametrizations
(\ref{eq:wcparametrization1}) and (\ref{eq:wcparametrization2}), one
obtains the twist-2 result for the reduced weak charged virtual
Compton scattering amplitude:
\begin{eqnarray}
\mathsf{\mathcal{T}}_{WC}^{\mu\nu} & = &
-\frac{1}{4\left(p\cdot q\right)}\left\{ \left[
\frac{1}{\left(p\cdot q_{2}\right)}
\left(p^{\mu}q_{2}^{\nu}+p^{\nu}q_{2}^{\mu}\right)-g^{\mu\nu}
\right]\right.\nonumber \\
&  & \times\left[\mathcal{H}_{WC}^{+}\left(\xi,t
\right)\bar{u}\left(p_{2},s_{2}\right)\not\! q_{2}u
\left(p_{1},s_{1}\right)+\mathcal{E}_{WC}^{+}\left(\xi,t\right)
\bar{u}\left(p_{2},s_{2}\right)
\frac{\left(\not\! q_{2}\not\! r-\not\! r\not\! q_{2}\right)}{4M}u
\left(p_{1},s_{1}\right)\right.\nonumber \\
&  & \hspace*{0.5cm}\left.-\widetilde{\mathcal{H}}_{WC}^{-}
\left(\xi,t\right)\bar{u}
\left(p_{2},s_{2}\right)\not\! q_{2}\gamma_{5}u\left(p_{1},s_{1}
\right)+\widetilde{\mathcal{E}}_{WC}^{-}\left(\xi,t\right)
\frac{\left(q_{2}\cdot r\right)}{2M}\bar{u}\left(p_{2},s_{2}
\right)\gamma_{5}u\left(p_{1},s_{1}\right)\right]\nonumber \\
&  & +\left[\frac{1}{\left(p\cdot q_{2}\right)}i
\epsilon^{\mu\nu\rho\eta}q_{2\rho}p_{\eta}\right]\nonumber \\
&  & \times\left[\widetilde{\mathcal{H}}_{WC}^{+}
\left(\xi,t\right)\bar{u}\left(p_{2},s_{2}\right)\not\! q_{2}
\gamma_{5}u\left(p_{1},s_{1}\right)-\widetilde{\mathcal{E}}_{WC}^{+}
\left(\xi,t\right)\frac{\left(q_{2}\cdot r\right)}{2M}
\bar{u}\left(p_{2},s_{2}\right)\gamma_{5}u\left(p_{1},s_{1}
\right)\right.\nonumber \\
&  & \hspace*{0.5cm}\left.-\mathcal{H}_{WC}^{\mathit{-}}
\left(\xi,t
\right)\bar{u}\left(p_{2},s_{2}\right)\not\! q_{2}u
\left(p_{1},s_{1}\right)-\mathcal{E}_{WC}^{\mathit{-}}
\left(\xi,t\right)\bar{u}\left(p_{2},s_{2}\right)
\frac{\left(\not\! q_{2}\not\! r-\not\! r\not\! q_{2}\right)}{4M}u
\left(p_{1},s_{1}\right)\right]\nonumber \\
&  & +\left[\frac{2}{\left(p\cdot q_{2}\right)}p^{\mu}p^{\nu}
\right]\nonumber \\
&  & \times\left[\mathcal{F}_{1}\left(t\right)\bar{u}
\left(p_{2},s_{2}\right)\not\! q_{2}u\left(p_{1},s_{1}\right)+
\mathcal{F}_{2}\left(t\right)\bar{u}\left(p_{2},s_{2}
\right)\frac{\left(\not\! q_{2}\not\! r-\not\! r
\not\! q_{2}\right)}{4M}u
\left(p_{1},s_{1}\right)\right.\nonumber \\
&  & \hspace*{0.5cm}\left.\left.-
\mathcal{G}_{A}\left(t\right)
\bar{u}\left(p_{2},s_{2}\right)\not\! q_{2}\gamma_{5}u
\left(p_{1},s_{1}\right)+\mathcal{G}_{P}\left(t\right)
\frac{\left(q_{2}\cdot r\right)}{2M}\bar{u}\left(p_{2},s_{2}
\right)\gamma_{5}u\left(p_{1},s_{1}\right)
\right]\right\} \;,
\label{eq:compactweakchargedamplitude}
\end{eqnarray}
where the integrals of OFPDs are now given by:
\begin{eqnarray}
\mathcal{H}_{WC}^{+\left(-\right)}\left(\xi,t\right) &
\equiv & \sum_{f,f'}\int_{-1}^{1}\frac{dx}{\left(x-\xi+i0\right)}
\left[Q_{+\left(-\right)}H_{f'f}^{+}\left(x,\xi,t
\right)+Q_{-\left(+\right)}H_{f'f}^{-}\left(x,\xi,t\right)
\right]\nonumber \\
& = & \sum_{f,f'}\int_{-1}^{1}dx\; H_{f'f}\left(x,
\xi,t\right)\left(\frac{Q_{f'}}{x-\xi+i0}\pm\frac{Q_{f}}{x+\xi-i0}
\right)\;,\label{eq:integralsofGPDsweakchargedcurrent1} \\
\mathcal{E}_{WC}^{+\left(-\right)}\left(\xi,t\right) &
\equiv & \sum_{f,f'}\int_{-1}^{1}\frac{dx}{\left(x-\xi+i0\right)}
\left[Q_{+\left(-\right)}E_{f'f}^{+}\left(x,\xi,t
\right)+Q_{-\left(+\right)}E_{f'f}^{-}
\left(x,\xi,t\right)\right]\nonumber \\
& = & \sum_{f,f'}\int_{-1}^{1}dx\; E_{f'f}
\left(x,\xi,t\right)\left(\frac{Q_{f'}}{x-\xi+i0}\pm
\frac{Q_{f}}{x+\xi-i0}\right)\;, \\
\widetilde{\mathcal{H}}_{WC}^{+\left(-\right)}\left(\xi,t\right) &
\equiv & \sum_{f,f'}\int_{-1}^{1}\frac{dx}{\left(x-\xi+i0\right)}
\left[Q_{+\left(-\right)}\widetilde{H}_{f'f}^{+}\left(x,\xi,t
\right)+Q_{-\left(+\right)}\widetilde{H}_{f'f}^{-}\left(x,\xi,t\right)
\right]\nonumber \\
& = & \sum_{f,f'}\int_{-1}^{1}dx\;\widetilde{H}_{f'f}
\left(x,\xi,t\right)\left(\frac{Q_{f'}}{x-\xi+i0}\mp
\frac{Q_{f}}{x+\xi-i0}\right)\;, \\
\widetilde{\mathcal{E}}_{WC}^{+\left(-\right)}\left(\xi,t\right) &
\equiv & \sum_{f,f'}\int_{-1}^{1}\frac{dx}{\left(x-\xi+i0\right)}
\left[Q_{+\left(-\right)}\widetilde{E}_{f'f}^{+}\left(x,\xi,t
\right)+Q_{-\left(+\right)}\widetilde{E}_{f'f}^{-}\left(x,\xi,t
\right)\right]\nonumber \\
& = & \sum_{f,f'}\int_{-1}^{1}dx\;\widetilde{E}_{f'f}
\left(x,\xi,t\right)\left(\frac{Q_{f'}}{x-\xi+i0}\mp
\frac{Q_{f}}{x+\xi-i0}\right)\;,
\label{eq:integralsofGPDsweakchargedcurrent4}
\end{eqnarray}
and
\begin{eqnarray}
\mathcal{F}_{1}\left(t\right) & \equiv &
\sum_{f,f'}Q_{-}\int_{-1}^{1}dx\; H_{f'f}^{-}
\left(x,\xi,t\right)=\sum_{f,f'}\left(Q_{f'}-Q_{f}\right)
\int_{-1}^{1}dx\; H_{f'f}\left(x,\xi,t\right)\ ,
\label{eq:formfactorsdefinition1} \\
\mathcal{F}_{2}\left(t\right) & \equiv &
\sum_{f,f'}Q_{-}\int_{-1}^{1}dx\; E_{f'f}^{-}
\left(x,\xi,t\right)=\sum_{f,f'}\left(Q_{f'}-Q_{f}
\right)\int_{-1}^{1}dx\; E_{f'f}
\left(x,\xi,t\right)\ , \\
\mathcal{G}_{A}\left(t\right) & \equiv &
\sum_{f,f'}Q_{-}\int_{-1}^{1}dx\;\widetilde{H}_{f'f}^{+}
\left(x,\xi,t\right)=\sum_{f,f'}\left(Q_{f'}-Q_{f}
\right)\int_{-1}^{1}dx\;
\widetilde{H}_{f'f}\left(x,\xi,t\right)\ , \\
\mathcal{G}_{P}\left(t\right) & \equiv &
\sum_{f,f'}Q_{-}\int_{-1}^{1}dx\;\widetilde{E}_{f'f}^{+}
\left(x,\xi,t\right)=\sum_{f,f'}\left(Q_{f'}-Q_{f}
\right)\int_{-1}^{1}dx\;
\widetilde{E}_{f'f}\left(x,\xi,t\right)\ .
\label{eq:formfactorsdefinition4}
\end{eqnarray}
It is important to note the difference in the structure between
the weak neutral and weak charged amplitudes. The latter has an
additional term, which is symmetric in indices $\mu$ and $\nu$
and is determined by the form factors
(\ref{eq:formfactorsdefinition1}) -- (\ref{eq:formfactorsdefinition4}).
This completes the derivation of the wDVCS amplitudes.

\section{Weak and Electromagnetic DVCS Processes\label{processes}}

In this section, we consider specific examples of wDVCS processes
for neutral and charged currents on an unpolarized nucleon target.
For the former, we examine both neutrino and electron (or charged
lepton in general) scattering, while the latter is illustrated using
the neutron-to-proton transition in neutrino scattering.
Before we examine the specific processes, however, we discuss the
kinematics which are common to all DVCS-like reactions.

\subsection{Kinematics}

\begin{figure}[H]
\begin{center}
\includegraphics[%
  scale=0.7]{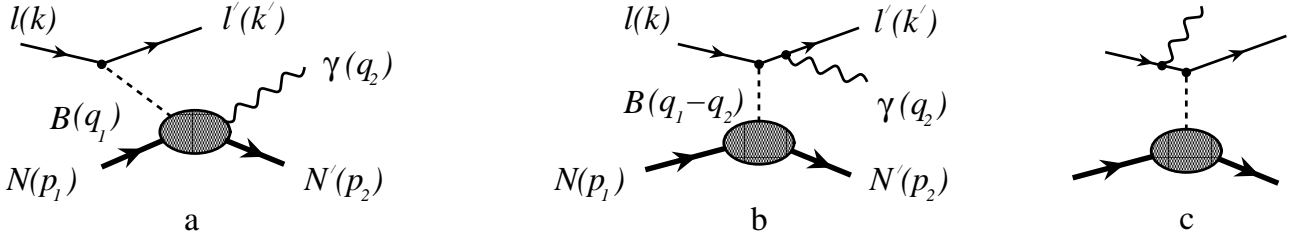}
\end{center}\caption{Weak DVCS (a) and Bethe-Heitler (b and c) diagrams
	contributing to the leptoproduction of a real photon.}
\label{weakprocess}
\end{figure}

The generalized DVCS process:
\begin{eqnarray}
l\left(k\right)+N\left(p_{1}\right) &
\longrightarrow & l'
\left(k'\right)+N'\left(p_{2}\right)+
\gamma\left(q_{2}\right)\ ,
\label{eq:reactionformula}
\end{eqnarray}
where a lepton $l$ (with four-momentum $k$) scatters
from a nucleon $N\left(p_{1}\right)$ to a final state
$l'\left(k'\right)$, nucleon $N'\left(p_{2}\right)$ and a real
photon $\gamma\left(q_{2}\right)$, is illustrated in
Fig.~\ref{weakprocess}. The wDVCS diagram, or Compton contribution,
is depicted in Fig.~\ref{weakprocess}(a), which corresponds to the
emission of a real photon from the nucleon ``blob''.
The other two diagrams, (b) and (c) in Fig.~\ref{weakprocess},
illustrate the Bethe-Heitler process, where the real photon is
emitted from either the initial or final lepton leg.
Here the nucleon ``blob'' represents the electroweak form factor,
while the upper part of each diagram can be calculated exactly in QED.

\begin{figure}[H]
\begin{center}
\includegraphics[%
  scale=0.70]{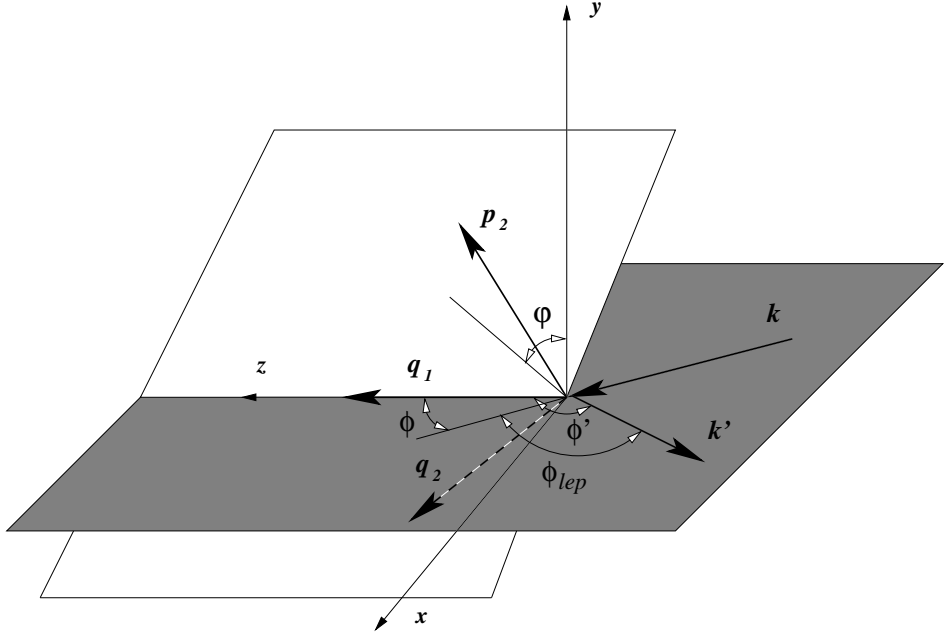}
\end{center}\caption{Kinematics of the generalized DVCS process in
	the target rest frame.}
\label{kinematics}
\end{figure}

We denote the various four-momenta in the target rest frame as
$k=\left(\omega,\vec{k}\right)$,
$p_{1}=\left(M,\vec{0}\right)$, $k'=\left(\omega',\vec{k}'\right)$,
$p_{2}=\left(E_{2},\vec{p}_{2}\right)$ and
$q_{2}=\left(\nu_{2},\vec{q}_{2}\right)$.
The differential cross section for lepton scattering from a nucleon
to a final state with a lepton, nucleon and a real photon is given by:
\begin{eqnarray}
d\sigma & = & \frac{1}{4\left(p_{1}\cdot k\right)}
\left|\mathrm{T}\right|^{2}
\frac{1}{\left(2\pi\right)^{5}}\;
\delta^{\left(4\right)}\left(k+p_{1}-k'-p_{2}-q_{2}\right)
\frac{d^{3}k'}{2\omega'}
\frac{d^{3}p_{2}}{2E_{2}}\frac{d^{3}q_{2}}{2\nu_{2}}\ ,
\label{eq:generaldiffcrosssection}
\end{eqnarray}
where T represents the invariant matrix element containing both
the Compton and Bethe-Heitler contributions:
\begin{eqnarray}
\mathrm{T} & = & \mathrm{T}_{C}+\mathrm{T}_{BH}\ .
\label{eq:totalamplitudeDVCS}
\end{eqnarray}
In the target rest frame, the total lepton--nucleon center of mass
energy squared is $s\equiv\left(p_{1}+k\right)^{2}=2M\omega+M^{2}$,
where we neglect lepton masses.
Integrating the cross section in Eq.~(\ref{eq:generaldiffcrosssection})
over the photon momentum yields:
\begin{eqnarray}
d\sigma & = & \frac{1}{4M\omega}\left|\mathrm{T}\right|^{2}
\frac{1}{\left(2\pi\right)^{5}}\;\delta
\left[\left(k+p_{1}-k'-p_{2}\right)^{2}\right]
\frac{\omega'd\omega'd\Omega'}{2}
\frac{d^{3}p_{2}}{2E_{2}}\ .
\label{eq:diffcrosssectionone}
\end{eqnarray}
The $\delta$-function here provides the constraint
$\hat{s}+M^{2}-2\left[\left(\nu_{1}+M\right)E_{2}-
\vec{q}_{1}\cdot\vec{p}_{2}\right]=0$,
where the invariant $\hat{s}\equiv\left(p_{1}+q_{1}\right)^{2}$ and
$q_{1}=k-k'=\left(\nu_{1},\vec{q}_{1}\right)$ is the four-momentum
of the virtual weak boson with the energy $\nu_{1}=\omega-\omega'$
and the magnitude of the three-momentum $\left|\vec{q}_{1}\right|=
\sqrt{\nu_{1}^{2}-q_{1}^{2}}$. In addition, we choose a coordinate
system (depicted in Fig.~\ref{kinematics}) where the virtual weak boson
four-momentum has no transverse components,
$q_{1}=\left(\nu_{1},0,0,\left|\vec{q}_{1}\right|\right)$,
and the incoming and outgoing lepton four-momenta are
$k=\omega\left(1,\sin\phi,0,\cos\phi\right)$
and $k'=\omega'\left(1,\sin\phi',0,\cos\phi'\right)$, respectively.
In this reference frame, the azimuthal angle of the recoil
nucleon corresponds to the angle $\varphi$ between the lepton and
nucleon scattering planes. Using the $\delta$-function in
Eq.~(\ref{eq:diffcrosssectionone}) to integrate over the polar angle
$\phi_{2}$ of the outgoing nucleon, and expressing its energy in terms of
the invariant momentum transfer $t\equiv\left(p_{1}-p_{2}\right)^{2}$ as
$E_{2}=M-t/\left(2M\right)$, we find:
\begin{eqnarray}
\cos\phi_{2} & = & -\frac{\left[1+\left(\omega-\omega'\right)/M
\right]t+q_{1}^{2}}{2\left|\vec{p}_{2}\right|\;
\sqrt{\left(\omega-\omega'\right)^{2}-q_{1}^{2}}}\,\,,
\label{eq:outgoinghadronapolarangle}
\end{eqnarray}
where
$\left|\vec{p}_{2}\right|=\sqrt{-t\left[1-t/\left(4M^{2}\right)\right]}$,
and the cross section takes the form:
\begin{eqnarray}
d\sigma & = & \frac{1}{32M\omega}\left|\mathrm{T}\right|^{2}
\frac{1}{\left(2\pi\right)^{4}}
\frac{\omega'}{\sqrt{\left(\omega-\omega'
\right)^{2}-q_{1}^{2}}}d\omega'dE_{2}d
\left(\cos\phi'\right)d\varphi\ .
\label{eq:diffcrosssection3DVCS}
\end{eqnarray}
Here the angle $\phi'$ denotes the polar angle of the
scattered lepton.

Instead of the kinematical variables ($\omega'$, $E_{2}$, $\cos\phi'$),
it is more convenient to express the differential cross section in
terms of the variables $\left(Q_{1}^{2},t,x_{B}\right)$,
where $Q_{1}^{2}\equiv-q_{1}^{2}$. In the target rest frame, we have
$Q_{1}^{2}=2M\omega yx_{B}$, where the invariant $y\equiv p_{1}
\cdot q_{1}/p_{1} \cdot k=\left(\omega-\omega'\right)/\omega$ and
accordingly, Eq.~(\ref{eq:outgoinghadronapolarangle}) turns into:
\begin{eqnarray}
\cos\phi_{2} & = & \frac{x_{B}M\left(1-t/Q_{1}^{2}\right)-t/
\left(2M\right)}{\left|\vec{p}_{2}\right|\;
\sqrt{1+4x_{B}^{2}M^{2}/Q_{1}^{2}}}\,\,.
\label{eq:outgoinghadronapolaranglenew}
\end{eqnarray}
Furthermore, from the conservation of three-momentum,
$\vec{q}_{1}=\vec{k}-\vec{k}'$, and writing the invariant
$Q_{1}^{2}=-\left(k-k'\right)^{2}=2\omega\omega'
\left(1-\cos\phi_{lep}\right)$, where $\phi_{lep}$ is the angle
between the incident and scattered lepton momenta, we get for the
polar angles of the incoming and scattered leptons:
\begin{eqnarray}
\cos\phi & = & \frac{1+2yx_{B}^{2}M^{2}/Q_{1}^{2}}{
\sqrt{1+4x_{B}^{2}M^{2}/Q_{1}^{2}}}
\label{eq:leptonpolarangles1}
\end{eqnarray}
and
\begin{eqnarray}
\cos\phi' & = & \frac{1}{\sqrt{1+4x_{B}^{2}M^{2}/Q_{1}^{2}}}
\left[1-\frac{2yx_{B}^{2}M^{2}}{\left(1-y\right)Q_{1}^{2}}
\right]\,\,,
\label{eq:leptonpolarangles2}
\end{eqnarray}
respectively. The differential cross section can then be written as:
\begin{eqnarray}
\frac{d^{4}\sigma}{dx_{B}dQ_{1}^{2}dtd\varphi} & = &
\frac{1}{32}\frac{1}{\left(2\pi\right)^{4}}
\frac{x_{B}y^{2}}{Q_{1}^{4}}
\frac{1}{\sqrt{1+4x_{B}^{2}M^{2}/Q_{1}^{2}}}
\left|{\textrm{T}}\right|^{2}\,\,.
\label{eq:generaldiffcrosswithQ1squared}
\end{eqnarray}
Finally, energy conservation, $M+\nu_{1}=E_{2}+\nu_{2}$, implies that
the energy of the outgoing real photon in the target rest frame is
$\nu_{2}=\left(Q_{1}^{2}+x_{B}t\right)/\left(2Mx_{B}\right)$.
Alternatively, we can write the invariant momentum transfer as
$t\equiv\left(q_{2}-q_{1}\right)^{2}=-Q_{1}^{2}-2\nu_{2}
\left(\omega y-\sqrt{\omega^{2}y^{2}+Q_{1}^{2}}\;
\cos\theta_{B\gamma}\right)$, where $\theta_{B\gamma}$ denotes the
scattering angle between the incoming virtual weak boson and
outgoing real photon. Combining both expressions gives:
\begin{eqnarray}
\cos\theta_{B\gamma} & = & \frac{Q_{1}^{2}\nu_{2}+x_{B}M
\left(Q_{1}^{2}+t\right)}{Q_{1}^{2}\nu_{2}
\sqrt{1+4x_{B}^{2}M^{2}/Q_{1}^{2}}}\,\,,
\label{eq:scatteringangleformula}
\end{eqnarray}
or, on the other hand, the invariant $t$ can be expressed as a function of
the angle $\theta_{B\gamma}$:
\begin{eqnarray}
t & = & -Q_{1}^{2}\,\frac{1+\left(1-\sqrt{1+4x_{B}^{2}M^{2}/Q_{1}^{2}}
\;\cos\theta_{B\gamma}\right)Q_{1}^{2}/\left(2x_{B}^{2}M^{2}
\right)}{1+\left(1-\sqrt{1+4x_{B}^{2}M^{2}/Q_{1}^{2}}\;
\cos\theta_{B\gamma}\right)Q_{1}^{2}/\left(2x_{B}M^{2}\right)}\,\,.
\label{eq:invariantt}
\end{eqnarray}

\begin{figure}[H]
\begin{center}
\includegraphics[%
  scale=0.75]{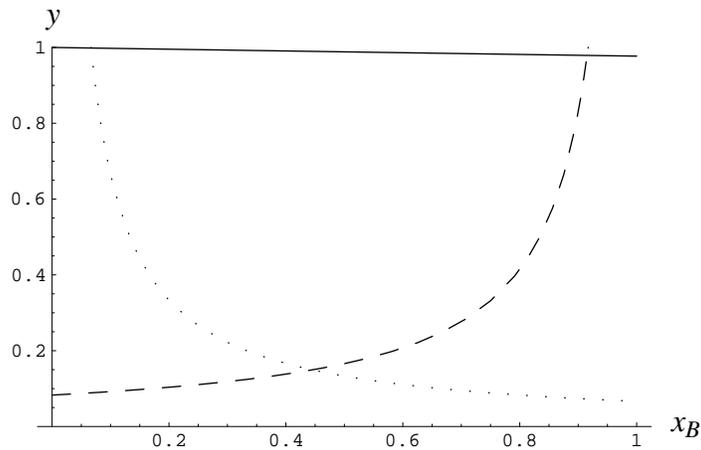}
\end{center}\caption{Kinematics of the DVCS process:
	the dashed curve gives $y_{\rm min}$ in Eq.~(\ref{eq:ymin}),
	the solid represents $y_{\rm max}$ in Eq.~(\ref{eq:ymax}),
	and the dotted corresponds to
	$y=Q_{\rm 1min}^{2}/\left[\left(s-M^{2}\right)x_{B}\right]$.
	The region bounded by the three curves is that kinematically
	allowed for $\hat{s}\geq4\;\mathrm{GeV}^{2}$ and
	$Q_{1}^{2}\geq2.5\;\mathrm{GeV}^{2}$ with an
	$\omega=20\;\mathrm{GeV}$ lepton beam.}
\label{xyplane}
\end{figure}

In the following, we determine the kinematically allowed region
for the generalized DVCS process, which requires finding the upper
and lower limits of the invariants $x_{B}$, $y$ and $t$.
The kinematics are subject to the following constraints:
\begin{enumerate}
\item	The energy of the incoming lepton beam is fixed at
	$\omega=20\;\mathrm{GeV}$.
\item	The invariant mass squared of the virtual boson--nucleon
	system should be above the resonance region,
	$\hat{s}\equiv\left(p_{1}+q_{1}\right)^{2}
	\geq\hat{s}_{\rm min}=4\;\mathrm{GeV}^{2}$.
\item	The virtuality of the incoming boson has to be large enough
	to ensure light-cone dominance of the scattering process,
	$Q_{1}^{2}\geq Q_{\rm 1min}^{2}=2.5\;\mathrm{GeV}^{2}$.
\item	The momentum transfer squared should be as small as possible,
	e.g. $0.1\;\mathrm{GeV^{2}}\leq\left|t\right|\leq0.2\;
	\mathrm{GeV^{2}}$, which yields a low-energy nucleon and a
	high-energy real photon with in the final state.
\end{enumerate}
The constraint (2) leads to a lower limit on the variable $y$:
\begin{eqnarray}
y_{\rm min} & = &
\frac{\hat{s}_{\rm min}-M^{2}}{\left(s-M^{2}\right)
\left(1-x_{B}\right)}\ .
\label{eq:ymin}
\end{eqnarray}
On the other hand, when the incoming lepton is
aligned along the $z$-axis, i.e. $\phi_{lep}=180^\circ$,
the scaling variable $y$ reaches its maximum value:
\begin{eqnarray}
y_{\rm max} & = & \left(1+\frac{M^{2}x_{B}}{s-M^{2}}
\right)^{-1}\ .
\label{eq:ymax}
\end{eqnarray}
The region in the $x_{B}y$ plane is then bounded by three curves,
illustrated in Fig.~\ref{xyplane}.
These are given by $y_{\rm min}$ in Eq.~(\ref{eq:ymin}), $y_{\rm max}$ in
Eq.~(\ref{eq:ymax}), and
$y=Q_{\rm 1min}^{2}/\left[\left(s-M^{2}\right)x_{B}\right]$,
which follows from the constraint (3).
Next, both the lower and upper limits of the invariant $t$ can be
determined in the virtual weak boson--nucleon center of mass frame.
Here the kinematic limits of $t$ are given by
(up to relative corrections of the order $x_{B}M^{2}/Q_{1}^{2}$):
\begin{eqnarray}
t_{\rm min}=\frac{-M^{2}x_{B}^{2}}{1-x_{B}
\left(1-M^{2}/Q_{1}^{2}\right)} &
\mathrm{and} & t_{\rm max}=
\frac{M^{2}x_{B}^{2}-2M^{2}x_{B}+Q_{1}^{2}
\left(x_{B}-1
\right)/x_{B}}{1-x_{B}
\left(1-M^{2}/Q_{1}^{2}\right)}\ ,
\label{eq:tlimits}
\end{eqnarray}
at the scattering angles of $0^\circ$ and $180^\circ$, respectively,
between the initial and final nucleons.

\begin{figure}[H]
\begin{center}
\includegraphics[%
  scale=0.75]{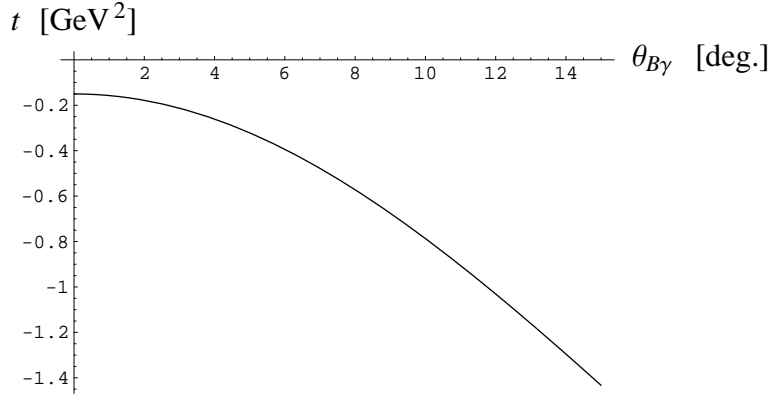}
\end{center}\caption{Invariant momentum transfer squared as a function
	of the scattering angle between the incoming virtual weak boson
	and outgoing real photon in the target rest frame,
	for $Q_{1}^{2}=2.5\;\mathrm{GeV}^{2}$ and $x_{B}=0.35$.}
\label{t}
\end{figure}

Finally, we select one kinematical point within the allowed region
in the $x_{B}y$ plane, namely $Q_{1}^{2}=2.5\; \mathrm{GeV}^{2}$
and $x_{B}=0.35$.
For this point, in Fig.~\ref{t} we plot the invariant momentum
transfer against the scattering angle $\theta_{B\gamma}$, which
varies between $-0.15\;\mathrm{GeV}^{2}$ and $-1.433\;\mathrm{GeV}^{2}$
for $0\leq\theta_{B\gamma}\leq15^\circ$. We also set the angle between
the lepton and nucleon scattering planes to $\varphi=0$. Then, with the
so-called in-plane kinematics, the polar angles of both
incoming and scattered leptons are fixed to $\phi=20.2^\circ$ and
$\phi'=25.3^\circ$, respectively.
%

\subsection{Model}
\label{sec:model}

To proceed with a numerical study of the wDVCS process, we need a
model of the OFPDs.
Since this is an exploratory study, rather than a detailed one intended
for comparison to experiment, we choose for simplicity a toy model to
illustrate the main features of wDVCS and the differences with
electromagnetic DVCS.

Inherent in our model are the following approximations:
\begin{itemize}
\item	The sea quark contributions are negligible, which implies
	that the ``plus'' OFPDs coincide with the ``minus'' OFPDs.
	For that reason, $H_{f}\equiv H_{f}^{val}$ and
	$\widetilde{H}_{f}\equiv\widetilde{H}_{f}^{val}$ for quark
	flavor $f=u$ or $d$, and similarly for the $E_{f}$ and
	$\widetilde{E}_{f}$ distributions.
\item	The $t$ dependence can be factorized from the other two
	scaling variables ($x$ and $\xi$) for all distributions,
	and the dependence of the OFPDs on $t$ is characterized
	by the corresponding form factors.
\item	The $\xi$ dependence of OFPDs appears only in the
	$\widetilde{E}_{f}$ distribution.
\end{itemize}
The parametrization of the unpolarized quark OFPDs in the proton is
taken from Ref.~\cite{Guichon:1998xv}:
\begin{eqnarray}
\label{eq:HandE1}
H_{u}^{val}\left(x,\xi,t\right) & = & u_{p}^{val}
\left(x\right)
F_{1u}\left(t\right)/2\ , \\
H_{d}^{val}\left(x,\xi,t\right) & = & d_{p}^{val}
\left(x\right)F_{1d}\left(t\right)\ , \\
E_{u}^{val}\left(x,\xi,t\right) & = & u_{p}^{val}
\left(x\right)F_{2u}\left(t\right)/2\ , \\
E_{d}^{val}\left(x,\xi,t\right) & = & d_{p}^{val}
\left(x\right)F_{2d}\left(t\right)\ ,
\label{eq:HandE2}
\end{eqnarray}
where the unpolarized valence quark distributions are given
by \cite{Radyushkin:1998rt}:
\begin{eqnarray}
u_{p}^{val}\left(x\right) & = & 1.89x^{-0.4}
\left(1-x\right)^{3.5}\left(1+6x\right)\ , \\
d_{p}^{val}\left(x\right) & = & 0.54x^{-0.6}
\left(1-x\right)^{4.2}\left(1+8x\right)\ .
\label{eq:unpolvalencedistributions}
\end{eqnarray}
They closely reproduce the corresponding GRV parametrizations
\cite{Gluck:1994uf} at a low normalization point,
$-q_{1}^{2}\simeq1\;\mathrm{GeV}^{2}$ \cite{Musatov:1999gv}.
The $u$- and $d$-quark form factors in
Eqs.~(\ref{eq:HandE1})--(\ref{eq:HandE2})
can be extracted from the proton and neutron Dirac and Pauli
form factors, neglecting strangeness and heavier flavors,
according to
$F_{1p\left(n\right)} = Q_{u}F_{1u\left(d\right)}
 		      + Q_{d}F_{1d\left(u\right)}$
and
$F_{2p\left(n\right)} = Q_{u}F_{2u\left(d\right)}
		      + Q_{d}F_{2d\left(u\right)}$,
which are in turn related to the Sachs electric and
magnetic form factors:
\begin{eqnarray}
F_{1p\left(n\right)}\left(t\right) & = &
\left[G_{Ep\left(n\right)}\left(t\right)-
\frac{t}{4M^{2}}G_{Mp\left(n\right)}
\left(t\right)\right]
\left(1-\frac{t}{4M^{2}}\right)^{-1}, \\
F_{2p\left(n\right)}\left(t\right) & = &
\left[G_{Mp\left(n\right)}
\left(t\right)-G_{Ep\left(n\right)}
\left(t\right)\right]\left(1-
\frac{t}{4M^{2}}\right)^{-1}.
\label{eq:f1andf2factors}
\end{eqnarray}
At small $t$, both the proton and neutron magnetic form factors,
as well as the proton electric form factor, are approximated by dipole
fits, while the electric neutron form factor is taken to be zero:
\begin{eqnarray}
G_{Ep}\left(t\right)=\frac{G_{Mp}
\left(t\right)}{1+\kappa_{p}}=
\frac{G_{Mn}\left(t\right)}{\kappa_{n}}=
\left(1-\frac{t}{\Lambda^{2}}
\right)^{-2},
& \mathrm{and} & G_{En}\left(t\right)=0\ ,
\label{eq:Sachsformfactors}
\end{eqnarray}
where $\kappa_{p}=1.793$ and $\kappa_{n}=-1.913$ are the proton and
neutron anomalous magnetic moments, respectively, and the parameter
$\Lambda^2=0.71\;\mathrm{GeV}^2$.

In the polarized case, we take for the valence distributions
the factorized {\em ansatz} from Ref.~\cite{Belitsky:2001ns}:
\begin{eqnarray}
\widetilde{H}_{u}^{val}\left(x,\xi,t\right) & = &
\Delta u_{p}^{val}
\left(x\right)\left(1-\frac{t}{m_{A}^{2}}
\right)^{-2}\ , \\
\widetilde{H}_{d}^{val}\left(x,\xi,t\right) & = &
\Delta d_{p}^{val}\left(x\right)
\left(1-\frac{t}{m_{A}^{2}}\right)^{-2}\ ,
\label{eq:Htilda}
\end{eqnarray}
where the mass $m_{A}=1.03\;\mathrm{GeV}$.
The polarized valence quark distributions in the proton can be
expressed in terms of the unpolarized distributions using a
parameterization motivated by the SU(6) quark model
\cite{Carlitz:1976in,Goshtasbpour:1995eh}:
\begin{eqnarray}
\Delta u_{p}^{val} & = &
\cos\theta_{D}\left(u_{p}^{val}-
\frac{2}{3}d_{p}^{val}\right)\ , \\
\Delta d_{p}^{val} & = & \cos\theta_{D}
\left(-\frac{1}{3}d_{p}^{val}\right)\ ,
\label{eq:polvalencedistributions}
\end{eqnarray}
where $\cos\theta_{D}=\left[1+0.06
\left(1-x^{2}\right)/\sqrt{x}\;\right]^{-1}$.
Finally, for the $\widetilde{E}_{f}$ distribution we use an {\em ansatz}
in which the distribution is dominated by the pion pole:
\begin{eqnarray}
\widetilde{E}_{u}^{val}\left(x,\xi,t\right) & = &
\frac{1}{2}F_{\pi}
\left(t\right)\frac{\theta
\left(\left|x\right|<\xi\right)}{2\xi}\;
\phi_{\pi}\left(\frac{x+\xi}{2\xi}
\right)\ , \\
\widetilde{E}_{d}^{val}\left(x,\xi,t\right)
& = & -\widetilde{E}_{u}^{val}
\left(x,\xi,t\right)\ .
\label{eq:Etilda}
\end{eqnarray}
The function $F_{\pi}\left(t\right)$ is assumed to have a form valid
for $-t\ll M^{2}$ \cite{Penttinen:1999th}:
\begin{eqnarray}
F_{\pi}\left(t\right) & = & \frac{4g_{A}M^{2}}{\left(m_{\pi}^{2}-t\right)}
\left[1-\frac{1.7\left(m_{\pi}^{2}-t\right)/
\mathrm{GeV^{2}}}{\left(1-t/2\;\mathrm{GeV^{2}}\right)^{2}}
\right],
\label{eq:Fpi}\end{eqnarray}
where $m_{\pi}\simeq 0.14\;\mathrm{GeV}$, and the axial charge of the
nucleon is $g_{A}\simeq 1.267$. For the pion distribution
amplitude in Eq.~(\ref{eq:Etilda}) we choose, for simplicity,
its asymptotic form:
\begin{eqnarray}
\phi_{\pi}\left(u\right) & = & 6u\left(1-u\right)\ .
\label{eq:Fipi}
\end{eqnarray}

Having defined the kinematics and described the model for the nucleon
OFPDs, in the rest of this section we examine several specific DVCS
processes.

\subsection{Weak neutral current scattering}

In this subsection we discuss two examples of the weak neutral current
scattering process, for neutrino--proton and electron--proton
scattering.
For the case of an unpolarized neutron target, one can use isospin
symmetry to express the neutron OFPDs in terms of proton OFPDs.
To make our comparisons meaningful, in all cases the relevant unpolarized
differential cross sections (\ref{eq:generaldiffcrosswithQ1squared}) will
be plotted as a function of the angle $\theta_{B\gamma}$ for the same
kinematical point, namely, $Q_{1}^{2}=2.5\;\mathrm{GeV}^{2}$ and
$x_{B}=0.35$, with a lepton beam energy $\omega=20\;\mathrm{GeV}$.

\subsubsection{Neutrino--proton scattering}

Since neutrinos do not interact with photons, neutrino scattering
from a proton via the exchange of a $Z^{0}$ boson measures the pure
Compton contribution, with no contribution from the Bethe-Heitler
process.
The T-matrix for this process,
\begin{eqnarray}
i\mathrm{T}_{\nu p} & = & \bar{u}\left(k'\right)
\left(\frac{-ig}{\cos\theta_{W}}
\right)\gamma^{\lambda}\left(\frac{c_{V}^{\nu}-
\gamma_{5}c_{A}^{\nu}}{2}
\right)u
\left(k\right)\left[
\frac{-i\left(g_{\nu\lambda}-q_{1\nu}q_{1\lambda}/M_{Z^{0}}^{2}
\right)}{q_{1}^{2}-M_{Z^{0}}^{2}}
\right]\left(\frac{-\left|e\right|g}{\cos\theta_{W}}
\right)\epsilon_{\mu}^{*}\left(q_{2}
\right)\left(-i\mathcal{T}_{WN}^{\mu\nu}\right)\ ,
\label{eq:Tmatrixneutrinoproton1}
\end{eqnarray}
is then solely given by the DVCS diagram, Fig.~\ref{weakprocess}(a).
The vector and axial vector couplings at the $\nu\nu Z^{0}$ vertex
are $c_{V}^{\nu}=c_{A}^{\nu}=1/2$.
By taking into account that $M_{Z^{0}}^{2}\gg Q_{1}^{2}$ and
$\cos\theta_{W}\equiv M_{W}/M_{Z^{0}}$, and further recalling that
$g^{2}/\left(8M_{W}^{2}\right)\equiv G_{F}/\sqrt{2}$,
where $G_{F}\simeq1.166\cdot10^{-5}\;\mathrm{GeV}^{-2}$ is
the Fermi coupling constant, we arrive at:
\begin{eqnarray}
\mathrm{T}_{\nu p} & = & \sqrt{2}\left|e\right|G_{F}\bar{u}
\left(k'\right)\gamma_{\nu}
\left(1-\gamma_{5}\right)u(k)\epsilon_{\mu}^{*}
\left(q_{2}\right)\mathcal{T}_{WN}^{\mu\nu}\ .
\label{eq:Tmatrixneutrinoproton2}
\end{eqnarray}
To compute the unpolarized cross section, one should average the
square of the amplitude $\mathrm{T}_{\nu p}$ over the spins of the
initial particles, and further sum it over the spins and polarizations
of the final particles. In particular, the summation over the photon
polarizations is performed using the Feynman gauge prescription, i.e.
one can replace
\begin{eqnarray}
\sum_{\gamma\; polar.}\epsilon_{\mu}^{*}
\left(q_{2}\right)\epsilon_{\alpha}
\left(q_{2}\right) & \longrightarrow & -g_{\mu\alpha}
\label{eq:feynmangauge}
\end{eqnarray}
by virtue of the Ward identity. As a result, we then write the
spin-averaged square of the T-matrix in terms of factorized neutrino
and weak neutral hadronic tensors,
\begin{eqnarray}
\overline{\left|\mathrm{T}_{\nu p}\right|^{2}} & = & 8
\pi\alpha G_{F}^{2}\ L_{\nu\beta}^{
\left(\nu\right)}H_{WN}^{\nu\beta}\ ,
\label{eq:Tmatrixsquaredneutrinoproton}
\end{eqnarray}
where $\alpha\equiv e^{2}/\left(4\pi\right)\simeq1/137$ is the
electromagnetic fine structure constant.
The neutrino tensor $L_{\nu\beta}^{\left(\nu\right)}$ is given by:
\begin{eqnarray}
L_{\nu\beta}^{\left(\nu\right)} & = & 8
\left[k_{\nu}k'_{\beta}+k_{\beta}k'_{\nu}-g_{\nu\beta}
\left(k\cdot k'\right)-i
\epsilon_{\nu\beta\sigma\tau}k^{\sigma}k'^{\tau}\right]\ .
\label{eq:neutrinotensorweakDVCS}
\end{eqnarray}
The weak neutral hadronic tensor $H_{WN}^{\nu\beta}$, on the other hand,
has a considerably more complicated structure.
In the DVCS kinematics, however, in which
$\mathcal{O}\left(t/q_{1}^{2}\right)$ and
$\mathcal{O}\left(M^{2}/q_{1}^{2}\right)$
terms are neglected, and using the Feynman gauge prescription,
this reduces to:
\begin{eqnarray}
H_{WN}^{\nu\beta} & = & -\frac{1}{2}
\mathcal{T}_{WN}^{\mu\nu}\left(
\mathcal{T}_{\mu WN}^{\beta}\right)^{*}\nonumber \\
& = & -\frac{1}{4}\left\{
\mathcal{C}_{1WN}\left[g^{\nu\beta}-
\frac{1}{\left(p\cdot q_{2}\right)}
\left(p^{\nu}q_{2}^{\beta}+p^{\beta}q_{2}^{\nu}
\right)+\frac{M^{2}}{\left(p\cdot q_{2}\right)^{2}}
\left(1-\frac{t}{4M^{2}}
\right)q_{2}^{\nu}q_{2}^{\beta}\right]-
\mathcal{C}_{2WN}\frac{1}{\left(p\cdot q_{2}\right)}i
\epsilon^{\nu\beta\delta\lambda}p_{\delta}q_{2\lambda}
\right\}\ , \nonumber \\
\label{eq:hadronictensorWN}
\end{eqnarray}
where the functions $\mathcal{C}_{1WN}$ and
$\mathcal{C}_{2WN}$ are given by:
\begin{eqnarray}
\mathcal{C}_{1WN} & = & \left(1-\xi^{2}\right)
\left(\left|\mathcal{H}_{WN}^{+}\right|^{2}+\left|
\mathcal{H}_{WN}^{-}\right|^{2}+\left|
\widetilde{\mathcal{H}}_{WN}^{+}\right|^{2}+\left|
\widetilde{\mathcal{H}}_{WN}^{-}\right|^{2}\right)-
\left(\xi^{2}+\frac{t}{4M^{2}}\right)\left(
\left|\mathcal{E}_{WN}^{+}\right|^{2}+\left|
\mathcal{E}_{WN}^{-}\right|^{2}\right)\nonumber \\
&  & \hspace*{0.5cm}-\xi^{2}\frac{t}{4M^{2}}
\left(\left|\widetilde{\mathcal{E}}_{WN}^{+}
\right|^{2}+\left|\widetilde{\mathcal{E}}_{WN}^{-}
\right|^{2}\right)-2\xi^{2}\Re\left(
\mathcal{H}_{WN}^{+*}\mathcal{E}_{WN}^{+}+
\mathcal{H}_{WN}^{-*}\mathcal{E}_{WN}^{-}+
\widetilde{\mathcal{H}}_{WN}^{+*}
\widetilde{\mathcal{E}}_{WN}^{+}+
\widetilde{\mathcal{H}}_{WN}^{-*}
\widetilde{\mathcal{E}}_{WN}^{-}\right), \\
\mathcal{C}_{2WN} & = & -2\left[\left(1-
\xi^{2}\right)\Re\left(\mathcal{H}_{WN}^{+*}
\mathcal{H}_{WN}^{-}+\widetilde{\mathcal{H}}_{WN}^{+*}
\widetilde{\mathcal{H}}_{WN}^{-}\right)-
\left(\xi^{2}+\frac{t}{4M^{2}}\right)\Re
\left(\mathcal{E}_{WN}^{+*}\mathcal{E}_{WN}^{-}
\right)-\xi^{2}\frac{t}{4M^{2}}\Re\left(
\widetilde{\mathcal{E}}_{WN}^{+*}
\widetilde{\mathcal{E}}_{WN}^{-}\right)\right.\nonumber \\
 &  & \hspace*{0.5cm}\left.-\xi^{2}\Re\left(
\mathcal{H}_{WN}^{+*}\mathcal{E}_{WN}^{-}+
\mathcal{E}_{WN}^{+*}\mathcal{H}_{WN}^{-}+
\widetilde{\mathcal{H}}_{WN}^{+*}
\widetilde{\mathcal{E}}_{WN}^{-}+
\widetilde{\mathcal{E}}_{WN}^{+*}
\widetilde{\mathcal{H}}_{WN}^{-}\right)\right] .
\label{eq:coefficientsweakneutralhadronictensor}
\end{eqnarray}
The result of the contraction of the tensor $H_{WN}^{\nu\beta}$
with $L_{\nu\beta}^{\left(\nu\right)}$ in
Eq.~(\ref{eq:neutrinotensorweakDVCS}) can be presented in a compact
form as:
\begin{eqnarray}
\overline{\left|\mathrm{T}_{\nu p}\right|^{2}} & = &
\frac{16\pi\alpha G_{F}^{2}Q_{1}^{2}}{y^{2}}\left\{
\left[1+\left(1-y\right)^{2}\right]\mathcal{C}_{1WN}-
\left[1-\left(1-y\right)^{2}\right]\mathcal{C}_{2WN}\right\} .
\label{eq:contractionweakneutralComptonneutrino}
\end{eqnarray}
%

\subsubsection{Electron--proton scattering}

Unlike in the neutrino case, for electron scattering both the Compton
and Bethe-Heitler processes contribute, in which case the $ep$
amplitude is given by the sum
$\mathrm{T}_{ep}=\mathrm{T}_{Cep}+\mathrm{T}_{BHep}$.
By replacing neutrinos with electrons in
Eq.~(\ref{eq:Tmatrixneutrinoproton1}), we can immediately write
for the Compton part of the T-matrix:
\begin{eqnarray}
\mathrm{T}_{Cep} & = & 2\sqrt{2}\left|e\right|G_{F}
\bar{u}\left(k'\right)\gamma_{\nu}
\left(c_{V}^{e}-\gamma_{5}c_{A}^{e}\right)u(k)
\epsilon_{\mu}^{*}\left(q_{2}
\right)\mathcal{T}_{WN}^{\mu\nu}\ ,
\label{eq:Tmatrixelectronproton}
\end{eqnarray}
where the couplings are now $c_{V}^{e}=-1/2+2\sin^{2}\theta_{W}$
and $c_{A}^{e}=-1/2$. The spin-averaged square of $\mathrm{T}_{Cep}$
then reads:
\begin{eqnarray}
\overline{\left|\mathrm{T}_{Cep}\right|^{2}} & = & 32
\pi\alpha G_{F}^{2}\ L_{\nu\beta}^{
\left(e\right)}H_{WN}^{\nu\beta}\ .
\label{eq:Tmatrixsquaredelectronproton}
\end{eqnarray}
The hadronic tensor $H_{WN}^{\nu\beta}$ is the same as in
Eq.~(\ref{eq:hadronictensorWN}), while the electron tensor has an
additional factor of $1/2$ from averaging over the initial electron
spins:
\begin{eqnarray}
L_{\nu\beta}^{\left(e\right)} & = & 2
\left\{ \left[\left(c_{V}^{e}
\right)^{2}+
\left(c_{A}^{e}\right)^{2}\right]
\left[k_{\nu}k'_{\beta}+k_{\beta}k'_{\nu}-g_{\nu\beta}
\left(k\cdot k'\right)\right]-2c_{V}^{e}c_{A}^{e}i
\epsilon_{\nu\beta\sigma\tau}k^{\sigma}k'^{\tau}
\right\}\ ,
\label{eq:leptonictensor1}
\end{eqnarray}
and hence
\begin{eqnarray}
\overline{\left|\mathrm{T}_{Cep}\right|^{2}} & = &
\frac{16\pi\alpha G_{F}^{2}Q_{1}^{2}}{y^{2}}\left\{
\left[\left(c_{V}^{e}\right)^{2}+\left(c_{A}^{e}
\right)^{2}\right]\left[1+\left(1-y\right)^{2}\right]
\mathcal{C}_{1WN}-2c_{V}^{e}c_{A}^{e}\left[1-
\left(1-y\right)^{2}\right]\mathcal{C}_{2WN}\right\} .
\label{eq:contractionweakneutralComptonelectron}
\end{eqnarray}

For the Bethe-Heitler contribution, since both the initial and
final leptons are electrons, both diagrams (b) and (c) in
Fig.~\ref{weakprocess} contribute.
The full Bethe-Heitler amplitude is given by:
\begin{eqnarray}
\mathrm{T}_{BHep} & = & 2\sqrt{2}\left|e\right|G_{F}
\epsilon_{\mu}^{*}\left(q_{2}\right)\bar{u}\left(k'\right)
\left[\frac{\gamma^{\mu}\left(\not\! k'+\not\! q_{2}\right)
\gamma^{\nu}\left(c_{V}^{e}-\gamma_{5}c_{A}^{e}
\right)}{\left(k'+q_{2}\right)^{2}}+\frac{\gamma^{\nu}
\left(c_{V}^{e}-\gamma_{5}c_{A}^{e}\right)
\left(\not\! k-\not\! q_{2}\right)\gamma^{\mu}}{\left(k-q_{2}
\right)^{2}}\right]u\left(k\right)\nonumber \\
&  & \times\left\langle p\left(p_{2},s_{2}\right)
\right|J_{\nu}^{NC}\left(0\right)\left|p\left(p_{1},s_{1}
\right)\right\rangle\ .
\label{eq:TmatrixwndvcsBHelectron}
\end{eqnarray}
Accordingly, the electron and hadronic tensors in the Bethe-Heitler
amplitude
\begin{eqnarray}
\overline{\left|\mathrm{T}_{BHep}\right|^{2}} & = & 32
\pi\alpha G_{F}^{2}\ L_{BH}^{\nu\beta}H_{\nu\beta}^{BH}
\label{eq:TmatrixsquaredBHelectronproton1}
\end{eqnarray}
are given by:
\begin{eqnarray}
L_{BH}^{\nu\beta} & = & -\frac{1}{2}\mathrm{Tr}
\left\{ \not\! k'\left[\frac{\left(\gamma^{\mu}
\not\! q_{2}+2k'^{\mu}\right)\gamma^{\nu}\left(c_{V}^{e}-
\gamma_{5}c_{A}^{e}\right)}{2\left(k'\cdot q_{2}\right)}+
\frac{\gamma^{\nu}\left(c_{V}^{e}-\gamma_{5}c_{A}^{e}
\right)\left(\not\! q_{2}\gamma^{\mu}-2k^{\mu}
\right)}{2\left(k\cdot q_{2}\right)}\right]
\right.\nonumber \\
&  & \hspace*{0.5cm}\left.\times\not\! k
\left[\frac{\gamma^{\beta}
\left(c_{V}^{e}-\gamma_{5}c_{A}^{e}\right)\left(
\not\! q_{2}\gamma_{\mu}+2k'_{\mu}\right)}{2\left(k'
\cdot q_{2}\right)}+\frac{\left(\gamma_{\mu}
\not\! q_{2}-2k_{\mu}\right)\gamma^{\beta}\left(c_{V}^{e}-
\gamma_{5}c_{A}^{e}\right)}{2\left(k\cdot q_{2}
\right)}\right]\right\} \nonumber \\
& = & \frac{2}{\left(k'\cdot q_{2}\right)}\left\{
\left[\left(c_{V}^{e}\right)^{2}+\left(c_{A}^{e}
\right)^{2}\right]\left[k^{\nu}q_{2}^{\beta}+k^{
\beta}q_{2}^{\nu}-g^{\nu\beta}\left(k\cdot q_{2}
\right)\right]-2c_{V}^{e}c_{A}^{e}i\epsilon^{\nu\beta
\sigma\tau}k_{\sigma}q{}_{2\tau}\right\} \nonumber \\
&  & +\frac{2}{\left(k\cdot q_{2}\right)}\left\{ \left[
\left(c_{V}^{e}\right)^{2}+\left(c_{A}^{e}\right)^{2}
\right]\left[k'^{\nu}q_{2}^{\beta}+k'^{
\beta}q_{2}^{\nu}-g^{\nu\beta}\left(k'\cdot q_{2}
\right)\right]+2c_{V}^{e}c_{A}^{e}i
\epsilon^{\nu\beta\sigma\tau}k'_{\sigma}q{}_{2\tau}
\right\} \nonumber \\
&  & +\frac{2}{\left(k'\cdot q_{2}\right)\left(k
\cdot q_{2}\right)}\left\{ \left[\left(c_{V}^{e}
\right)^{2}+\left(c_{A}^{e}\right)^{2}\right]
\left[\left(k\cdot q_{2}\right)\left[k^{\nu}k'^{
\beta}+k^{\beta}k'^{\nu}+2k'^{\nu}k'^{\beta}\right]-
\left(k'\cdot q_{2}\right)\left[k^{\nu}k'^{\beta}+k^{
\beta}k'^{\nu}+2k^{\nu}k^{\beta}\right]
\right.\right.\nonumber \\
&  & \hspace*{0.5cm}\left.+\left(k\cdot k'\right)
\left[k^{\nu}q_{2}^{\beta}+k^{\beta}q_{2}^{\nu}-k'^{
\nu}q_{2}^{\beta}-k'^{\beta}q_{2}^{\nu}+2k^{\nu}k'^{
\beta}+2k^{\beta}k'^{\nu}\right]+2g^{\nu\beta}
\left(k\cdot k'\right)\left[\left(k'\cdot q_{2}
\right)-\left(k\cdot q_{2}\right)-\left(k\cdot k'
\right)\right]\right]\nonumber \\
&  & \hspace*{0.5cm}-2c_{V}^{e}c_{A}^{e}i
\epsilon^{\nu\beta\sigma\tau}
\left[\left(k\cdot k'\right)\left[2k{}_{\sigma}k'{}_{
\tau}+k{}_{\sigma}q{}_{2\tau}+k'{}_{\sigma}q{}_{2\tau}
\right]+\left[\left(k\cdot q_{2}\right)-
\left(k'\cdot q_{2}\right)
\right]k{}_{\sigma}k'{}_{\tau}\right]\bigg\rbrace
\label{eq:BHleptonicWNelectron}
\end{eqnarray}
and
\begin{eqnarray}
H_{\nu\beta}^{BH} & = & \frac{1}{2}\sum_{s_{1},s_{2}}
\left\langle p\left(p_{2},s_{2}\right)
\right|J_{\nu}^{NC}\left(0\right)\left|p
\left(p_{1},s_{1}\right)\right\rangle \left
\langle p\left(p_{2},s_{2}\right)\right|J_{\beta}^{NC}
\left(0\right)\left|p\left(p_{1},s_{1}
\right)\right\rangle ^{*}\ ,
\label{eq:BHhadronicWNelectron1}
\end{eqnarray}
respectively.
The weak neutral transition current $J_{\nu}^{NC}$ in
Eq.~(\ref{eq:BHhadronicWNelectron1}) consists of vector
and axial vector parts:
\begin{eqnarray}
J_{\nu}^{NC}\left(0\right) & = & \frac{1}{2}
\left[V_{\nu}^{NC}\left(0\right)-A_{\nu}^{NC}
\left(0\right)\right]\nonumber \\
& = & \frac{1}{2}\sum_{f}\left[c_{V}^{f}\bar{\psi}_{f}
\left(0\right)\gamma_{\nu}
\psi_{f}\left(0\right)-c_{A}^{f}\bar{\psi}_{f}
\left(0\right)\gamma_{\nu}\gamma_{5}\psi_{f}
\left(0\right)\right]\ .
\label{eq:hadronictransitioncurrent}
\end{eqnarray}
Their matrix elements are parametrized as
(see e.g. Ref.~\cite{Thomas:2001kw}):
\begin{eqnarray}
\left\langle p\left(p_{2},s_{2}\right)
\right|V_{\nu}^{NC}\left(0\right)\left|p\left(p_{1},s_{1}
\right)\right\rangle  & = & \bar{u}\left(p_{2},s_{2}
\right)\left[F_{1}^{NC}\left(t\right)
\gamma_{\nu}-F_{2}^{NC}\left(t\right)
\frac{i\sigma_{\nu\lambda}r^{\lambda}}{2M}
\right]u\left(p_{1},s_{1}\right)\ , \\
\left\langle p\left(p_{2},s_{2}\right)
\right|A_{\nu}^{NC}\left(0\right)
\left|p\left(p_{1},s_{1}\right)\right
\rangle  & = & \bar{u}\left(p_{2},s_{2}\right)
\left[G_{A}^{NC}\left(t\right)\gamma_{\nu}
\gamma_{5}-G_{P}^{NC}\left(t\right)
\frac{\gamma_{5}r_{\nu}}{2M}\right]u
\left(p_{1},s_{1}\right)\ ,
\label{eq:transitioncurrentparametrization}
\end{eqnarray}
which then gives the following expression for the
hadronic tensor:
\begin{eqnarray}
H_{\nu\beta}^{BH} & = & \frac{1}{4}\left\{ t\left[g_{\nu\beta}-
\frac{r_{\nu}r_{\beta}}{t}\right]\left[F_{1}^{NC}\left(t
\right)+F_{2}^{NC}\left(t\right)\right]^{2}+4\left[p_{1\nu}-
\frac{r_{\nu}}{2}\right]\left[p_{1\beta}-\frac{r_{\beta}}{2}
\right]\left[\left(F_{1}^{NC}\left(t\right)\right)^{2}-
\frac{t}{4M^{2}}\left(F_{2}^{NC}\left(t\right)\right)^{2}
\right]\right\} \nonumber \\
&  & +\left[\left(p_{1\nu}-\frac{r_{\nu}}{2}
\right)\left(p_{1\beta}-\frac{r_{\beta}}{2}
\right)-M^{2}\left(1-\frac{t}{4M^{2}}\right)g_{\nu\beta}
\right]\left(G_{A}^{NC}\left(t\right)\right)^{2}+
\frac{r_{\nu}r_{\beta}}{4}\left(1-\frac{t}{4M^{2}}
\right)\left(G_{P}^{NC}\left(t\right)\right)^{2} \nonumber \\
&  & -\frac{r_{\nu}r_{\beta}}{4}\left[G_{A}^{NC}
\left(t\right)+G_{P}^{NC}\left(t\right)\right]^{2}-i
\epsilon_{\nu\beta\sigma\tau}p_{1}^{\sigma}p_{2}^{\tau}
\left[F_{1}^{NC}\left(t\right)+F_{2}^{NC}\left(t\right)
\right]G_{A}^{NC}\left(t\right)\ .
\label{eq:BHhadronicWNelectron2}
\end{eqnarray}
We express the result of the contraction of
Eq.~(\ref{eq:BHhadronicWNelectron2}) with the leptonic tensor in
Eq.~(\ref{eq:BHleptonicWNelectron}) in terms of the kinematical
invariants $x_{B}$, $y$, $Q_{1}^{2}$ and $t$, and the scalar product:
\begin{eqnarray}
k\cdot q_{2} & = & \omega\nu_{2}\left[1-
\left(\sin\phi\sin\theta_{B\gamma}+\cos\phi\cos
\theta_{B\gamma}\right)\right]\nonumber \\
& = & Q_{1}^{2}\left.\frac{1+x_{B}t/Q_{1}^{2}}{y
\left(1+4x_{B}^{2}M^{2}/Q_{1}^{2}\right)}\right\{ 1-
\frac{y}{2}-\frac{1}{2}\left(1+
\frac{2yx_{B}^{2}M^{2}}{Q_{1}^{2}}\right)\left(
\frac{1+t/Q_{1}^{2}}{1+x_{B}t/Q_{1}^{2}}
\right)\nonumber \\
&  & \left.-\sqrt{\left(1-y-
\frac{y^{2}x_{B}^{2}M^{2}}{Q_{1}^{2}}\right)
\left[1-\frac{1+t/Q_{1}^{2}}{1+x_{B}t/Q_{1}^{2}}-
\frac{x_{B}^{2}M^{2}}{Q_{1}^{2}}\left(
\frac{1+t/Q_{1}^{2}}{1+x_{B}t/Q_{1}^{2}}
\right)^{2}\right]}\;\right\} .
\label{eq:kdotq2}
\end{eqnarray}
Moreover, it is convenient to introduce the dimensionless
variables $\tau\equiv t/Q_{1}^{2}$,
$\mu\equiv M^{2}/Q_{1}^{2}$ and
$\kappa\equiv\left(k\cdot q_{2}\right)/Q_{1}^{2}$. The latter
can be written in the following form:
\begin{eqnarray}
\kappa & = & \frac{1}{2y\left(1+4x_{B}^{2}\mu\right)}
\left\{ 1-2\mathcal{K}-\tau\left[1-x_{B}\left(2-y
\right)+2x_{B}^{2}y\mu\right]+2x_{B}^{2}y\mu\right\} -
\frac{1}{2}\ ,
\label{eq:kappa}
\end{eqnarray}
where
\begin{eqnarray}
\mathcal{K} & = & \sqrt{-\left(1-y-x_{B}^{2}y^{2}
\mu\right)\left[\left(1-x_{B}\right)\tau+x_{B}
\left(1-x_{B}\right)\tau^{2}+x_{B}^{2}y^{2}
\mu\left(1+\tau\right)^{2}\right]}
\label{eq:kinematicalfactor}
\end{eqnarray}
is the $1/\sqrt{Q_{1}^{2}}$-power suppressed kinematical factor.
In terms of these variables, the Bethe-Heitler squared amplitude
can then be written:
\begin{eqnarray}
\overline{\left|\mathrm{T}_{BHep}\right|^{2}} & = & -32\pi
\alpha G_{F}^{2}Q_{1}^{2}\left[\frac{1}{2\kappa
\left(2\kappa+\tau+1
\right)x_{B}^{2}y^{2}}\right]\nonumber \\
&  & \times\Bigg\lbrace\left[\left(c_{V}^{e}\right)^{2}+
\left(c_{A}^{e}\right)^{2}\right]\left[x_{B}^{2}y^{2}\tau
\left[\left(2\kappa+1\right)^{2}+\left(2\kappa+\tau\right)^{2}
\right]\left[F_{1}^{NC}\left(t\right)+F_{2}^{NC}\left(t
\right)\right]^{2}\right.\nonumber \\
&  & +2\left[2\tau\left(1-y\right)-x_{B}y\tau
\left(4\kappa+\tau+1\right)+y^{2}\left[\tau+x_{B}\tau\left(2
\kappa+\tau\right)+x_{B}^{2}\mu\left[1+8\kappa^{2}+\tau^{2}+4
\kappa\left(1+\tau\right)\right]\right]\right]\nonumber \\
&  & \left.\times\left[\left(F_{1}^{NC}\left(t\right)
\right)^{2}-\frac{t}{4M^{2}}\left(F_{2}^{NC}\left(t\right)
\right)^{2}\right]\right]\nonumber \\
&  & +\left[4\tau-2y\tau\left[2+x_{B}\left(4
\kappa+\tau+1\right)\right]+y^{2}\left[2\tau+2x_{B}\tau\left(2
\kappa+\tau\right)+x_{B}^{2}\left(\tau-2\mu\right)\left[1+8
\kappa^{2}+\tau^{2}+4\kappa\left(1+\tau\right)\right]
\right]\right]\nonumber \\
&  & \times\left[\left(c_{V}^{e}\right)^{2}+\left(c_{A}^{e}
\right)^{2}\right]\left(G_{A}^{NC}\left(t\right)
\right)^{2}\nonumber \\
&  & +4x_{B}y\tau\left[2\left(1-\tau\right)+y\left[2\left(2
\kappa+\tau\right)-x_{B}\left(1-\tau\right)\left(4\kappa+
\tau+1\right)\right]\right]c_{V}^{e}c_{A}^{e}\left[F_{1}^{NC}
\left(t\right)+F_{2}^{NC}\left(t\right)\right]G_{A}^{NC}
\left(t\right)\Bigg\rbrace,
\label{eq:TmatrixsquaredBHelectronproton2}
\end{eqnarray}
where the term $1/\left[2\kappa\left(2\kappa+\tau+1
\right)\right]$ comes from the lepton propagators.
Note that, in contrast to the Compton contribution,
Eq.~(\ref{eq:TmatrixsquaredBHelectronproton2}) is the result
of the exact calculation. Each of the vector form factors can be
further decomposed into linear combinations of flavor triplet,
octet and singlet form factors:
\begin{eqnarray}
F_{1\left(2\right)}^{NC}\left(t\right) & = &
\left(1-2\sin^{2}\theta_{W}\right)
\left[F_{1\left(2\right)}^{3}\left(t\right)+
\frac{1}{6}F_{1\left(2\right)}^{8}\left(t\right)
\right]-\frac{1}{6}F_{1\left(2\right)}^{0}\left(t\right)\ ,
\label{eq:F12formfactors}
\end{eqnarray}
while the axial and pseudoscalar form factors can be written,
in general, as differences between the isovector and strangeness
form factors:
\begin{eqnarray}
G_{A\left(P\right)}^{NC}
\left(t\right) & = & G_{A\left(P\right)}^{3}
\left(t\right)-
\frac{1}{2}G_{A\left(P\right)}^{s}\left(t\right)\ .
\label{eq:GAGPformfactors}
\end{eqnarray}
If one further neglects sea quark contributions, then these
form factors can be written as:
\begin{eqnarray}
F_{1\left(2\right)}^{3}\left(t\right) & = &
\frac{1}{2}\left[F_{1\left(2\right)u}
\left(t\right)-F_{1\left(2\right)d}
\left(t\right)\right]\ , \\
F_{1\left(2\right)}^{8}\left(t
\right)=F_{1\left(2\right)}^{0}\left(t\right) & = &
\left[F_{1\left(2\right)u}\left(t
\right)+F_{1\left(2\right)d}
\left(t\right)\right]\ , \\
G_{A\left(P\right)}^{NC}
\left(t\right) & = & G_{A\left(P\right)}^{3}
\left(t\right).
\label{eq:modelFFs}
\end{eqnarray}
Recall that the $t$ dependence of the $u$- and $d$-quark
form factors is given by the Dirac and Pauli form factors,
whereas for the axial and pseudoscalar form factors we use the
parametrizations:
\begin{eqnarray}
\label{eq:axialandpseudoscalar1}
G_{A}^{NC}\left(t\right) & = &
\frac{g_{A}}{2}\left(1-
\frac{t}{m_{A}^{2}}\right)^{-2}\ , \\
G_{P}^{NC}\left(t\right) & = &
\frac{G_{A}^{NC}\left(t\right)}{2}
\left( \frac{4M^{2}}{m_{\pi}^{2}-t}\ \right).
\label{eq:axialandpseudoscalar2}
\end{eqnarray}

The interference terms between the Compton and Bethe-Heitler
contributions,
\begin{eqnarray}
\mathcal{I}_{ep} & = & \mathrm{T}_{Cep}
\mathrm{T}_{BHep}^{*}+
\mathrm{T}_{Cep}^{*}
\mathrm{T}{}_{BHep}\ ,
\label{eq:interferencewn1}
\end{eqnarray}
are particularly interesting, since they are linear in the
integrals of OFPDs. Substituting the expressions in
Eqs.~(\ref{eq:Tmatrixelectronproton}) and
(\ref{eq:TmatrixwndvcsBHelectron}) into
Eq.~(\ref{eq:interferencewn1}), and averaging and summing
over the initial and final spins, respectively,
the interference term can be written in terms of the
electron and hadronic traces as:
\begin{eqnarray}
\mathcal{I}_{ep} & = & -4\pi\alpha G_{F}^{2}\left\{
\mathrm{Tr}\left\{ \not\! k'\gamma_{\nu}\left(c_{V}^{e}-
\gamma_{5}c_{A}^{e}\right)\not\! k\left[\frac{
\gamma^{\beta}\left(c_{V}^{e}-\gamma_{5}c_{A}^{e}\right)
\left(\not\! q_{2}\gamma_{\mu}+2k'_{\mu}\right)}{2\left(k'
\cdot q_{2}\right)}+\frac{\left(\gamma_{\mu}
\not\! q_{2}-2k_{\mu}\right)\gamma^{\beta}\left(c_{V}^{e}-
\gamma_{5}c_{A}^{e}\right)}{2\left(k\cdot q_{2}\right)}
\right]\right\} \right.\nonumber \\
&  & \times\mathrm{Tr}\left\{ \left(\not\! p_{2}+M
\right)\hat{\mathcal{T}}_{WN}^{\mu\nu}\left(
\not\! p_{1}+M\right)\left[F_{1}^{NC}\left(t\right)
\gamma_{\beta}+F_{2}^{NC}\left(t\right)\frac{i
\sigma_{\beta\tau}r^{\tau}}{2M}-G_{A}^{NC}\left(t
\right)\gamma_{\beta}\gamma_{5}-G_{P}^{NC}\left(t
\right)\frac{\gamma_{5}r_{\beta}}{2M}\right]\right\}
\nonumber\\
&  & +\mathrm{Tr}\left\{ \not\! k\gamma_{\beta}
\left(c_{V}^{e}-\gamma_{5}c_{A}^{e}\right)\not\! k'
\left[\frac{\left(\gamma^{\mu}\not\! q_{2}+2k'^{\mu}
\right)\gamma^{\nu}\left(c_{V}^{e}-\gamma_{5}c_{A}^{e}
\right)}{2\left(k'\cdot q_{2}\right)}+\frac{
\gamma^{\nu}\left(c_{V}^{e}-\gamma_{5}c_{A}^{e}\right)
\left(\not\! q_{2}\gamma^{\mu}-2k^{\mu}\right)}{2
\left(k\cdot q_{2}\right)}\right]\right\} \nonumber \\
&  & \left.\times\mathrm{Tr}\left\{ \left(
\not\! p_{1}+M\right)\left(\hat{\mathcal{T}}_{
\mu WN}^{\beta}\right)^{*}\left(\not\! p_{2}+M
\right)\left[F_{1}^{NC}\left(t\right)\gamma_{
\nu}-F_{2}^{NC}\left(t\right)\frac{i
\sigma_{\nu\lambda}r^{\lambda}}{2M}-G_{A}^{NC}
\left(t\right)\gamma_{\nu}\gamma_{5}+G_{P}^{NC}
\left(t\right)\frac{\gamma_{5}r_{\nu}}{2M}
\right]\right\} \right\}\ , \nonumber \\
\label{eq:interferencewn3}
\end{eqnarray}
where the amplitude ${\hat{\mathcal{T}}}_{WN}^{\mu\nu}$ is the
spinorless part of the reduced virtual Compton scattering amplitude
$\mathcal{T}_{WN}^{\mu\nu}$:
\begin{eqnarray}
\mathbf{\mathcal{T}}_{WN}^{\mu\nu} & = & \bar{u}\left(p_{2},s_{2}
\right){\hat{\mathcal{T}}}_{WN}^{\mu\nu} u\left(p_{1},s_{1}
\right).
\label{eq:amplitudenospinors}
\end{eqnarray}
To   be consistent, we need to  keep    the same    level   of    accuracy 
as in the Compton part,
i.e. we should neglect terms  that   are of $\mathcal{O}\left(t/q_{1}^{2}\right)$ and
$\mathcal{O}\left(M^{2}/q_{1}^{2}\right)$ order. Accordingly, the variable
$\kappa$ becomes
$\kappa=\left[\left(1-y\right)-2\mathcal{K}\right]/\left(2y\right)$.
Furthermore,   we  should only keep terms linear in $\mathcal{K}$.  After
the contraction, we obtain:
\begin{eqnarray}
\mathcal{I}_{ep} & = & -16\pi\alpha G_{F}^{2}Q_{1}^{2}\left[
\frac{1}{2\kappa\left(2\kappa+\tau+1\right)x_{B}y^{3}}\right]
\left\{ \left[\left(c_{V}^{e}\right)^{2}+\left(c_{A}^{e}
\right)^{2}\right]\left[4\mathcal{K}\left(2-2y+y^{2}
\right)\right]\right.\nonumber \\
&  & \times\left[F_{1}^{NC}\left(t\right)
\Re\left(\mathcal{H}_{WN}^{+}
\right)-\frac{t}{4M^{2}}F_{2}^{NC}\left(t\right)\Re\left(
\mathcal{E}_{WN}^{+}\right)+G_{A}^{NC}\left(t\right)\Re\left(
\widetilde{\mathcal{H}}_{WN}^{-}\right)\right.\nonumber \\
&  & \left.+\xi\left[\left[F_{1}^{NC}\left(t\right)+F_{2}^{NC}
\left(t\right)\right]\Re\left(\widetilde{\mathcal{H}}_{WN}^{+}
\right)+G_{A}^{NC}\left(t\right)
\left[\Re\left(\mathcal{H}_{WN}^{-}
\right)+\Re\left(\mathcal{E}_{WN}^{-}\right)\right]
\right]\right]\nonumber \\
&  & +c_{V}^{e}c_{A}^{e}\left[8\mathcal{K}y\left(2-y\right)
\right]\left[F_{1}^{NC}\left(t\right)\Re\left(\mathcal{H}_{WN}^{-}
\right)-\frac{t}{4M^{2}}F_{2}^{NC}\left(t\right)\Re\left(
\mathcal{E}_{WN}^{-}\right)+G_{A}^{NC}\left(t\right)\Re\left(
\widetilde{\mathcal{H}}_{WN}^{+}\right)\right.\nonumber \\
&  & \left.\left.+\xi\left[\left[F_{1}^{NC}
\left(t\right)+F_{2}^{NC}\left(t\right)\right]
\Re\left(\widetilde{\mathcal{H}}_{WN}^{-}\right)+G_{A}^{NC}
\left(t\right)\left[\Re\left(\mathcal{H}_{WN}^{+}\right)+
\Re\left(\mathcal{E}_{WN}^{+}\right)\right]\right]\right]
\right\}\ .
\label{eq:interferencewn4}
\end{eqnarray}
%

\subsection{Weak charged current scattering}

In the weak charged current sector, we consider neutrino scattering
from a neutron via the exchange of a $W^{+}$ boson, producing a proton
in the final state.
The T-matrix of the Compton contribution in this case is:
\begin{eqnarray}
\mathrm{T}_{C\nu n} & = & \sqrt{2}\left|e\right|G_{F}\bar{u}
\left(k'\right)\gamma_{\nu}
\left(1-\gamma_{5}\right)u\left(k\right)\epsilon_{\mu}^{*}
\left(q_{2}\right)
\mathcal{T}_{WC}^{\mu\nu}\ ,
\label{eq:TmatrixCWC}
\end{eqnarray}
where the amplitude $\mathcal{T}_{WC}^{\mu\nu}$ is,
in general, given by Eq.~(\ref{eq:compactweakchargedamplitude}).
In our simple model, however, the quark flavors $f$ and $f'$ in
Eqs.~(\ref{eq:integralsofGPDsweakchargedcurrent1})
-- (\ref{eq:integralsofGPDsweakchargedcurrent4}) are $d$ and $u$,
respectively, and the coefficients equal to $Q_{+}=1/6$ and
$Q_{-}=1/2$.
To proceed, we relate the flavor nondiagonal OFPDs to flavor
diagonal ones using isospin symmetry.
Specifically, the nucleon matrix elements
$\left\langle p\left(p_{2},s_{2}\right)\right|
\mathcal{O}^{ud\pm}\left(z\left|0\right.\right)
\left|n\left(p_{1},s_{1}\right)\right\rangle$
and
$\left\langle p\left(p_{2},s_{2}\right)\right|
\mathcal{O}_{5}^{ud\pm}\left(z\left|0\right.\right)
\left|n\left(p_{1},s_{1}\right)\right\rangle$
in Eqs.~(\ref{eq:contractedoperatorsWC1}) and
(\ref{eq:contractedoperatorsWC2}) that are nondiagonal in quark
flavor are expressed in terms of the flavor diagonal matrix elements
according to Ref.~\cite{Mankiewicz:1997aa}:
\begin{eqnarray}
\left\langle p\left(p_{2},s_{2}\right)\right|\mathcal{O}^{ud\pm}
\left(z\left|0\right.\right)\left|n
\left(p_{1},s_{1}\right)\right\rangle  & = & \left
\langle p\left(p_{2},s_{2}\right)
\right|\mathcal{O}^{u\pm}\left(z\left|0\right.\right)
\left|p\left(p_{1},s_{1}
\right)\right\rangle-\left\langle p\left(p_{2},s_{2}\right)
\right|\mathcal{O}^{d\pm}
\left(z\left|0\right.\right)\left|p\left(p_{1},s_{1}
\right)\right\rangle ,
\label{eq:isospinsymmetry}
\end{eqnarray}
and similarly for $\left\langle p\left(p_{2},s_{2}\right)
\right|\mathcal{O}_{5}^{ud\pm}\left(z\left|0\right.\right)
\left|n\left(p_{1},s_{1}\right)\right\rangle$.
The convolution integrals in
Eqs.~(\ref{eq:integralsofGPDsweakchargedcurrent1}) --
(\ref{eq:integralsofGPDsweakchargedcurrent4}) and
Eqs.~(\ref{eq:formfactorsdefinition1}) --
(\ref{eq:formfactorsdefinition4}) then become:
\begin{eqnarray}
\label{eq:integralsofGPDsweakchargedcurrentmode1}
\mathcal{H}_{WC}^{+\left(-\right)}\left(\xi,t\right) & = &
\int_{-1}^{1}dx\;\left[H_{u}\left(x,\xi,t\right)-H_{d}
\left(x,\xi,t\right)\right]\left(\frac{Q_{u}}{x-\xi+i0}
\pm\frac{Q_{d}}{x+\xi-i0}\right)\;,\\
\mathcal{E}_{WC}^{+\left(-\right)}\left(\xi,t\right) & = &
\int_{-1}^{1}dx\;\left[E_{u}\left(x,\xi,t\right)-E_{d}
\left(x,\xi,t\right)\right]\left(\frac{Q_{u}}{x-\xi+i0}\pm
\frac{Q_{d}}{x+\xi-i0}\right)\;,\\
\widetilde{\mathcal{H}}_{WC}^{+\left(-\right)}
\left(\xi,t\right) & = & \int_{-1}^{1}dx\;\left[
\widetilde{H}_{u}\left(x,\xi,t\right)-\widetilde{H}_{d}
\left(x,\xi,t\right)\right]\left(\frac{Q_{u}}{x-\xi+i0}
\mp\frac{Q_{d}}{x+\xi-i0}\right)\;,\\
\widetilde{\mathcal{E}}_{WC}^{+\left(-\right)}
\left(\xi,t\right) & = & \int_{-1}^{1}dx\;
\left[\widetilde{E}_{u}\left(x,\xi,t\right)-
\widetilde{E}_{d}\left(x,\xi,t\right)\right]
\left(\frac{Q_{u}}{x-\xi+i0}\mp\frac{Q_{d}}{x+\xi-i0}
\right)\;,
\label{eq:integralsofGPDsweakchargedcurrentmode4}
\end{eqnarray}
and
\begin{eqnarray}
\label{eq:formfactorsdefinition1mode}
\mathcal{F}_{1}\left(t\right) & = &
\left(Q_{u}-Q_{d}\right)\left[F_{1u}
\left(t\right)-F_{1d}\left(t\right)\right]\;,\\
\mathcal{F}_{2}\left(t\right) & = &
\left(Q_{u}-Q_{d}\right)\left[F_{2u}
\left(t\right)-F_{2d}\left(t\right)\right]\;,\\
\mathcal{G}_{A}\left(t\right) & = &
\left(Q_{u}-Q_{d}\right)g_{A}\left(t\right)\;,\\
\mathcal{G}_{P}\left(t\right) & = &
\left(Q_{u}-Q_{d}\right)g_{P}\left(t\right)\;.
\label{eq:formfactorsdefinition4mode}
\end{eqnarray}
The spin-averaged square of the T-matrix in Eq.~(\ref{eq:TmatrixCWC})
can then be written as:
\begin{eqnarray}
\overline{\left|\mathrm{T}_{C\nu n}\right|^{2}} & = & 8
\pi\alpha G_{F}^{2}\ L_{\nu\beta}^{
\left(\mu\right)}H_{WC}^{\nu\beta}\ ,
\label{eq:Tmatrixsquaredneutrinoneutron}
\end{eqnarray}
where the weak charged hadronic tensor is given by:
\begin{eqnarray}
H_{WC}^{\nu\beta} & = & -\frac{1}{2}
\mathcal{T}_{WC}^{\mu\nu}
\left(\mathcal{T}_{\mu WC}^{\beta}
\right)^{*}\nonumber \\
& = & -\frac{1}{4}\left\{
\mathcal{C}_{1WC}\left[g^{\nu\beta}-
\frac{1}{\left(p\cdot q_{2}\right)}
\left(p^{\nu}q_{2}^{\beta}+p^{\nu}q_{2}^{\beta}
\right)+\frac{M^{2}}{\left(p\cdot q_{2}\right)^{2}}
\left(1-\frac{t}{4M^{2}}
\right)q_{2}^{\nu}q_{2}^{\beta}\right]-
\mathcal{C}_{2WC}\frac{1}{\left(p\cdot q_{2}
\right)}i\epsilon^{\nu\beta\delta\lambda}p_{
\delta}q_{2\lambda}\right.\nonumber \\
&  & \left.+\frac{2M^{2}}{
\left(p\cdot q_{2}\right)^{2}}\left(1-
\frac{t}{4M^{2}}\right)\left[
\mathcal{C}_{3WC}\ p^{\nu}q_{2}^{\beta}+
\mathcal{C}_{4WC}\ p^{\beta}q_{2}^{\nu}+2
\mathcal{C}_{5WC}\ p^{\nu}p^{\beta}
\right]\right\} ,
\label{eq:hadronictensorWC}
\end{eqnarray}
with the following set of functions:
\begin{eqnarray}
\mathcal{C}_{1WC} & = & \left(1-\xi^{2}\right)\left(\left|
\mathcal{H}_{WC}^{+}\right|^{2}+\left|\mathcal{H}_{WC}^{-}
\right|^{2}+\left|\widetilde{\mathcal{H}}_{WC}^{+}\right|^{2}+
\left|\widetilde{\mathcal{H}}_{WC}^{-}\right|^{2}\right)-
\left(\xi^{2}+\frac{t}{4M^{2}}\right)\left(\left|
\mathcal{E}_{WC}^{+}\right|^{2}+\left|\mathcal{E}_{WC}^{-}
\right|^{2}\right)\nonumber \\
&  & -\xi^{2}\frac{t}{4M^{2}}\left(\left|
\widetilde{\mathcal{E}}_{WC}^{+}\right|^{2}+\left|
\widetilde{\mathcal{E}}_{WC}^{-}\right|^{2}\right)-2
\xi^{2}\Re\left(\mathcal{H}_{WC}^{+*}\mathcal{E}_{WC}^{+}+
\mathcal{H}_{WC}^{-*}\mathcal{E}_{WC}^{-}+
\widetilde{\mathcal{H}}_{WC}^{+*}\widetilde{\mathcal{E}}_{WC}^{+}+
\widetilde{\mathcal{H}}_{WC}^{-*}
\widetilde{\mathcal{E}}_{WC}^{-}\right),\\
\mathcal{C}_{2WC} & = & -2\left[\left(1-\xi^{2}\right)
\Re\left(\mathcal{H}_{WC}^{+*}\mathcal{H}_{WC}^{-}+
\widetilde{\mathcal{H}}_{WC}^{+*}\widetilde{\mathcal{H}}_{WC}^{-}
\right)-\left(\xi^{2}+\frac{t}{4M^{2}}\right)\Re\left(
\mathcal{E}_{WC}^{+*}\mathcal{E}_{WC}^{-}\right)-\xi^{2}
\frac{t}{4M^{2}}\Re\left(\widetilde{\mathcal{E}}_{WC}^{+*}
\widetilde{\mathcal{E}}_{WC}^{-}\right)\right.\nonumber \\
&  & -\xi^{2}\Re\left(\mathcal{H}_{WC}^{+*}
\mathcal{E}_{WC}^{-}+\mathcal{E}_{WC}^{+*}
\mathcal{H}_{WC}^{-}+\widetilde{\mathcal{H}}_{WC}^{+*}
\widetilde{\mathcal{E}}_{WC}^{-}+\widetilde{\mathcal{E}}_{WC}^{+*}
\widetilde{\mathcal{H}}_{WC}^{-}\right)\bigg],\\
\mathcal{C}_{3WC} & = & \left(1-\xi^{2}\right)\left(
\mathcal{F}_{1}\mathcal{H}_{WC}^{+*}+\mathcal{G}_{A}
\widetilde{\mathcal{H}}_{WC}^{-*}\right)-\left(\xi^{2}+
\frac{t}{4M^{2}}\right)\left(\mathcal{F}_{2}
\mathcal{E}_{WC}^{+*}\right)\nonumber \\
&  & -\xi^{2}\frac{t}{4M^{2}}\left(\mathcal{G}_{P}
\widetilde{\mathcal{E}}_{WC}^{-*}\right)-\xi^{2}\left(
\mathcal{F}_{1}\mathcal{E}_{WC}^{+*}+\mathcal{F}_{2}
\mathcal{H}_{WC}^{+*}+\mathcal{G}_{A}
\widetilde{\mathcal{E}}_{WC}^{-*}+\mathcal{G}_{P}
\widetilde{\mathcal{H}}_{WC}^{-*}\right),\\
\mathcal{C}_{4WC} & = & \left(1-\xi^{2}\right)
\left(\mathcal{F}_{1}\mathcal{H}_{WC}^{+}+
\mathcal{G}_{A}\widetilde{\mathcal{H}}_{WC}^{-}\right)-
\left(\xi^{2}+\frac{t}{4M^{2}}\right)\left(\mathcal{F}_{2}
\mathcal{E}_{WC}^{+}\right)\nonumber \\
&  & -\xi^{2}\frac{t}{4M^{2}}\left(\mathcal{G}_{P}
\widetilde{\mathcal{E}}_{WC}^{-}\right)-\xi^{2}\left(
\mathcal{F}_{1}\mathcal{E}_{WC}^{+}+\mathcal{F}_{2}
\mathcal{H}_{WC}^{+}+\mathcal{G}_{A}
\widetilde{\mathcal{E}}_{WC}^{-}+\mathcal{G}_{P}
\widetilde{\mathcal{H}}_{WC}^{-}\right), \\
\mathcal{C}_{5WC} & = & \left(1-\xi^{2}
\right)\left(\mathcal{F}_{1}^{2}+
\mathcal{G}_{A}^{2}\right)-\left(\xi^{2}+
\frac{t}{4M^{2}}\right)\mathcal{F}_{2}^{2}-\xi^{2}
\frac{t}{4M^{2}}\mathcal{G}_{P}^{2}-2\xi^{2}\left(
\mathcal{F}_{1}\mathcal{F}_{2}+\mathcal{G}_{A}
\mathcal{G}_{P}\right).
\label{eq:coefficientsweakchargedhadronictensor}
\end{eqnarray}
As opposed to the weak neutral case, the additional functions
$\mathcal{C}_{3WC}$, $\mathcal{C}_{4WC}$ and $\mathcal{C}_{5WC}$
appear as a result of the extra term in the weak charged
amplitude (\ref{eq:compactweakchargedamplitude}).
Moreover, by neglecting the mass of the outgoing charged lepton,
the leptonic tensor $L_{\nu\beta}^{\left(\mu\right)}$ coincides
with the neutrino tensor $L_{\nu\beta}^{\left(\nu\right)}$ in
Eq.~(\ref{eq:neutrinotensorweakDVCS}). Consequently, after
ignoring terms $\mathcal{O}\left(t/q_{1}^{2}\right)$ and
$\mathcal{O}\left(M^{2}/q_{1}^{2}\right)$,
Eq.~(\ref{eq:Tmatrixsquaredneutrinoneutron}) simplifies into:
\begin{eqnarray}
\overline{\left|\mathrm{T}_{C\nu n}\right|^{2}} & = &
\frac{16\pi\alpha G_{F}^{2}Q_{1}^{2}}{y^{2}}\left\{
\left[1+\left(1-y\right)^{2}\right]\mathcal{C}_{1WC}-
\left[1-\left(1-y\right)^{2}\right]\mathcal{C}_{2WC}
\right\}.
\label{eq:contractionweakchargedCompton}
\end{eqnarray}

For the Bethe-Heitler background, only diagram (b) of
Fig.~\ref{weakprocess} contributes, for which the amplitude
is given by:
\begin{eqnarray}
\mathrm{T}_{BH\nu n} & = & \sqrt{2}\left|e\right|G_{F}
\epsilon_{\mu}^{*}\left(q_{2}\right)
\bar{u}\left(k'\right)\left[\frac{\gamma^{\mu}
\left(\not\! k'+\not\! q_{2}\right)\gamma^{\nu}
\left(1-\gamma_{5}\right)}{\left(k'+q_{2}\right)^{2}}
\right]u\left(k\right)\left
\langle p\left(p_{2},s_{2}\right)
\right|J_{\nu}^{CC}\left(0\right)
\left|n\left(p_{1},s_{1}\right)\right\rangle\ .
\label{eq:TmatrixBHcharged}
\end{eqnarray}
The spin-averaged square of the amplitude $\mathrm{T}_{BH\nu n}$
can then be written as:
\begin{eqnarray}
\overline{\left|\mathrm{T}_{BH\nu n}\right|^{2}} & = & 8
\pi\alpha G_{F}^{2}\ L_{BH}^{\nu\beta}H_{\nu\beta}^{BH}\ ,
\label{eq:TmatrixsquaredBHcharged1}
\end{eqnarray}
with the leptonic and hadronic tensors given by:
\begin{eqnarray}
L_{BH}^{\nu\beta} & = & -\mathrm{Tr}\left\{ \not\! k'
\left[\frac{\left(\gamma^{\mu}\not\! q_{2}+2k'^{\mu}
\right)\gamma^{\nu}\left(1-\gamma_{5}\right)}{2
\left(k'\cdot q_{2}\right)}\right]\right.\left.
\not\! k\left[\frac{\gamma^{\beta}\left(1-\gamma_{5}
\right)\left(\not\! q_{2}\gamma_{\mu}+2k'_{\mu}
\right)}{2\left(k'\cdot q_{2}\right)}
\right]\right\} \nonumber \\
& = & \frac{8}{\left(k'\cdot q_{2}\right)}
\left[k^{\nu}q_{2}^{\beta}+k^{\beta}q_{2}^{
\nu}-g^{\nu\beta}\left(k\cdot q_{2}\right)-i
\epsilon^{\nu\beta\sigma\tau}k_{\sigma}q{}_{2\tau}
\right],
\label{eq:wcdvcsBHleptonictensor1}\\
H_{\nu\beta}^{BH} & = & \frac{1}{2}\sum_{s_{1},s_{2}}
\left\langle p\left(p_{2},s_{2}\right)\right|J_{\nu}^{CC}
\left(0\right)\left|n\left(p_{1},s_{1}\right)
\right\rangle \left\langle p\left(p_{2},s_{2}
\right)\right|J_{\beta}^{CC}\left(0\right)
\left|n\left(p_{1},s_{1}\right)
\right\rangle ^{*} ,
\label{eq:wcdvcsBHleptonictensor2}
\end{eqnarray}
respectively. The matrix element of the weak charged
transition current between the nucleon states is defined as:
\begin{eqnarray}
\left\langle p\left(p_{2},s_{2}\right)\right|J_{\nu}^{CC}
\left(0\right)\left|n\left(p_{1},s_{1}
\right)\right\rangle  & = & \left\langle p
\left(p_{2},s_{2}\right)\right|\bar{\psi}_{p}
\left(0\right)\gamma_{\nu}\frac{1}{2}\left(1-
\gamma_{5}\right)\psi_{n}\left(0\right)
\left|n\left(p_{1},s_{1}\right)
\right\rangle\ .
\label{eq:transitioncurrentcc}
\end{eqnarray}
Using the isospin symmetry relation in Eq.~(\ref{eq:isospinsymmetry})
between the flavor nondiagonal and diagonal nucleon matrix elements,
we can parametrize the vector part of the matrix element in
Eq.~(\ref{eq:transitioncurrentcc}) in terms of the Dirac and Pauli
form factors for each quark flavor:
\begin{eqnarray}
\left\langle p\left(p_{2},s_{2}\right)\right|\bar{\psi}_{p}
\left(0\right)\gamma_{\nu}\psi_{n}\left(0\right)
\left|n\left(p_{1},s_{1}\right)\right\rangle  & = &
\left\langle p\left(p_{2},s_{2}\right)\right|\bar{\psi}_{u}
\left(0\right)\gamma_{\nu}\psi_{u}\left(0\right)\left|p
\left(p_{1},s_{1}\right)\right\rangle -\left\langle p
\left(p_{2},s_{2}\right)\right|\bar{\psi}_{d}\left(0\right)
\gamma_{\nu}\psi_{d}\left(0\right)\left|p\left(p_{1},s_{1}
\right)\right\rangle \nonumber \\
& = & \bar{u}\left(p_{2},s_{2}\right)\left\{ \left[F_{1u}
\left(t\right)-F_{1d}\left(t\right)\right]\gamma_{\nu}-
\left[F_{2u}\left(t\right)-F_{2d}\left(t\right)\right]
\frac{i\sigma_{\nu\lambda}r^{\lambda}}{2M}\right\} u
\left(p_{1},s_{1}\right)\ . \nonumber \\
\label{eq:vectorpart}
\end{eqnarray}
In addition, for the axial vector current part we have:
\begin{eqnarray}
\left\langle p\left(p_{2},s_{2}\right)\right|
\bar{\psi}_{p}\left(0\right)
\gamma_{\nu}\gamma_{5}\psi_{n}\left(0\right)
\left|n\left(p_{1},s_{1}\right)
\right\rangle  & = & \bar{u}\left(p_{2},s_{2}\right)
\Big[g_{A}\left(t\right)\gamma_{\nu}\gamma_{5}
\left.-g_{P}\left(t\right)
\frac{\gamma_{5}r_{\nu}}{2M}\right]u\left(p_{1},s_{1}\right)\ ,
\label{eq:axialvectorpart}
\end{eqnarray}
where the $t$ dependence of the axial and pseudoscalar from factors
follows the parametrizations in Eqs.~(\ref{eq:axialandpseudoscalar1})
and (\ref{eq:axialandpseudoscalar2}):
\begin{eqnarray}
\label{eq:axialandpseudoscalarWC1}
g_{A}\left(t\right) & = & g_{A}
\left(1-\frac{t}{m_{A}^{2}}
\right)^{-2}\ , \\
g_{P}\left(t\right) & = & g_{A}\left(t\right)
\left( \frac{4M^{2}}{m_{\pi}^{2}-t} \right) .
\label{eq:axialandpseudoscalarWC2}
\end{eqnarray}
The hadronic tensor here is given by Eq.~(\ref{eq:BHhadronicWNelectron2}),
with the form factors corresponding to the weak neutral transition current
replaced by those describing the weak charged current interaction.
The squared amplitude in Eq.~(\ref{eq:TmatrixsquaredBHcharged1}) can
then be written as:
\begin{eqnarray}
\overline{\left|\mathrm{T}_{BH\nu n}\right|^{2}} & = & -8\pi
\alpha G_{F}^{2}Q_{1}^{2}\left[\frac{2}{\left(2\kappa+
\tau+1\right)x_{B}^{2}y^{2}}
\right]\nonumber \\
&  & \times\left\{ x_{B}^{2}y^{2}\left[2\kappa+1+
\tau\left(4\kappa+1\right)\right]\left[\left[F_{1u}\left(t
\right)-F_{1d}\left(t\right)\right]+\left[F_{2u}\left(t
\right)-F_{2d}\left(t\right)\right]\right]^{2}\right.\nonumber \\
&  & +y\left[2x_{B}\left(1-\tau\right)-4+x_{B}y
\left[2-x_{B}\left(1-\tau\right)+\kappa\left[4-2x_{B}
\left(1-4\mu\right)\right]\right]\right]\nonumber \\
&  & \times\left[\left[F_{1u}\left(t\right)-F_{1d}
\left(t\right)\right]^{2}-\frac{t}{4M^{2}}\left[F_{2u}
\left(t\right)-F_{2d}\left(t\right)
\right]^{2}\right]\nonumber \\
&  & +y\left[2x_{B}\left(1-\tau\right)-4+x_{B}y
\left[2-x_{B}\left(1-\tau\right)+\kappa\left[4-2x_{B}
\left(1+4\mu\right)+4x_{B}\tau\right]\right]
\right]g_{A}^{2}\left(t\right)\nonumber \\
&  & +x_{B}^{2}y^{2}\left(2\kappa+\tau+1
\right)\left[\left[g_{A}\left(t\right)+g_{P}\left(t\right)
\right]^{2}-\left[1-\frac{t}{4M^{2}}\right]g_{P}^{2}
\left(t\right)\right]\nonumber \\
&  & \left.+4x_{B}y\left[1+\tau-y\left(2
\kappa+1\right)\left(1+x_{B}\tau
\right)\right]\left[\left[F_{1u}\left(t\right)-F_{1d}
\left(t\right)\right]+\left[F_{2u}\left(t\right)-F_{2d}
\left(t\right)\right]\right]g_{A}\left(t\right)\right\} \ .
\label{eq:TmatrixsquaredBHcharged2}
\end{eqnarray}

Finally, the interference contribution between the Compton and
Bethe-Heitler amplitudes has the following structure:
\begin{eqnarray}
\mathcal{I}_{\nu n} & = & -2\pi\alpha G_{F}^{2}\left\{ \mathrm{Tr}
\left\{ \not\! k'\gamma_{\nu}\left(1-\gamma_{5}\right)
\not\! k\left[\frac{\gamma^{\beta}\left(1-\gamma_{5}\right)
\left(\not\! q_{2}\gamma_{\mu}+2k'_{\mu}\right)}{2\left(k'
\cdot q_{2}\right)}\right]\right\} \mathrm{Tr}\Bigg\lbrace
\left(\not\! p_{2}+M\right)\hat{\mathcal{T}}_{WC}^{\mu\nu}
\left(\not\! p_{1}+M\right)\right.\nonumber \\
&  & \times\left[\left[F_{1u}\left(t\right)-F_{1d}\left(t
\right)\right]\gamma_{\beta}+\left[F_{2u}\left(t\right)-F_{2d}
\left(t\right)\right]\frac{i\sigma_{\beta\tau}r^{\tau}}{2M}-g_{A}
\left(t\right)\gamma_{\beta}\gamma_{5}-g_{P}\left(t\right)
\frac{\gamma_{5}r_{\beta}}{2M}\right]\Bigg\rbrace\nonumber \\
&  & +\mathrm{Tr}\left\{ \not\! k\gamma_{\beta}\left(1-
\gamma_{5}\right)\not\! k'\left[\frac{\left(\gamma^{\mu}
\not\! q_{2}+2k'^{\mu}\right)\gamma^{\nu}\left(1-\gamma_{5}
\right)}{2\left(k'\cdot q_{2}\right)}\right]\right\}
\mathrm{Tr}\Bigg\lbrace\left(\not\! p_{1}+M\right)\left(
\hat{\mathcal{T}}_{\mu WC}^{\beta}\right)^{*}\left(
\not\! p_{2}+M\right)\nonumber \\
&  & \left.\times\left[\left[F_{1u}\left(t\right)-F_{1d}
\left(t\right)\right]\gamma_{\nu}-\left[F_{2u}
\left(t\right)-F_{2d}\left(t\right)\right]
\frac{i\sigma_{\nu\lambda}r^{\lambda}}{2M}-g_{A}
\left(t\right)\gamma_{\nu}\gamma_{5}+g_{P}\left(t\right)
\frac{\gamma_{5}r_{\nu}}{2M}
\right]\Bigg\rbrace\right\}
\label{eq:interferencewc1}
\end{eqnarray}
in analogy with Eq.~(\ref{eq:interferencewn3}), or more explicitly, after
performing the contractions:
\begin{eqnarray}
\mathcal{I}_{\nu n} & = & -32\pi\alpha G_{F}^{2}Q_{1}^{2}
\left[\frac{1}{2\kappa\left(2\kappa+\tau+1
\right)x_{B}y^{3}}\right]\nonumber \\
&  & \times\left\{ \frac{1}{y}\left[-\left(2-x_{B}\right)
\left(1-y\right)\left[2-y\left(2-y\right)\right]+2\mathcal{K}
\left[8-x_{B}\left(2-y\right)^{3}-y\left[11-y\left(6-y
\right)\right]\right]\right]\right.\nonumber \\
&  & \times\left[\left[F_{1u}\left(t\right)-F_{1d}\left(t
\right)\right]\Re\left(\mathcal{H}_{WC}^{+}\right)-
\frac{t}{4M^{2}}\left[F_{2u}\left(t\right)-F_{2d}
\left(t\right)\right]\Re\left(\mathcal{E}_{WC}^{+}
\right)+g_{A}\left(t\right)\Re\left(
\widetilde{\mathcal{H}}_{WC}^{-}\right)\right]\nonumber \\
&  & +\xi\left[-\frac{x_{B}}{\left(2-x_{B}\right)y}
\left[-\left(2-x_{B}\right)\left(1-y\right)\left[2-y
\left(2-y\right)\right]-2\mathcal{K}\left(2-y\right)
\left[x_{B}\left(2-y\right)^{2}-2\left[3-y\left(3-y
\right)\right]\right]\right]\right]\nonumber \\
&  & \times\left[\left[\left[F_{1u}\left(t\right)-F_{1d}
\left(t\right)\right]+\left[F_{2u}\left(t\right)-F_{2d}
\left(t\right)\right]\right]\left[\Re\left(
\mathcal{H}_{WC}^{+}\right)+\Re\left(\mathcal{E}_{WC}^{+}
\right)\right]+g_{A}\left(t\right)\left[
\Re\left(\widetilde{\mathcal{H}}_{WC}^{-}\right)+\Re\left(
\widetilde{\mathcal{E}}_{WC}^{-}\right)
\right]\right.\nonumber \\
&  & \left.+g_{P}\left(t\right)\left[\Re\left(
\widetilde{\mathcal{H}}_{WC}^{-}\right)+\frac{t}{4M^{2}}
\Re\left(\widetilde{\mathcal{E}}_{WC}^{-}\right)
\right]\right]\nonumber \\
&  & +\xi\left[-2\mathcal{K}\left(1-y\right)^{2}
\right]\left[\left[\left[F_{1u}\left(t\right)-F_{1d}
\left(t\right)\right]+\left[F_{2u}\left(t\right)-F_{2d}
\left(t\right)\right]\right]\Re\left(\widetilde{
\mathcal{H}}_{WC}^{-}\right)+g_{A}\left(t\right)
\left[\Re\left(\mathcal{H}_{WC}^{+}\right)+\Re\left(
\mathcal{E}_{WC}^{+}\right)
\right]\right]\nonumber \\
&  & +\left[-\left(2-x_{B}\right)\left(2-y
\right)\left(1-y\right)-2\mathcal{K}\left[-7+y
\left(6-y\right)+x_{B}\left[6-y\left(6-y\right)
\right]\right]\right]\nonumber \\
&  & \times\left[\left[F_{1u}\left(t\right)-F_{1d}
\left(t\right)\right]\Re\left(\mathcal{H}_{WC}^{-}
\right)-\frac{t}{4M^{2}}\left[F_{2u}\left(t\right)-F_{2d}
\left(t\right)\right]\Re\left(\mathcal{E}_{WC}^{-}
\right)+g_{A}\left(t\right)\Re\left(
\widetilde{\mathcal{H}}_{WC}^{+}\right)\right]\nonumber \\
&  & +\xi\left[-\frac{x_{B}}{\left(2-x_{B}\right)}
\left[-\left(2-x_{B}\right)\left(2-y\right)
\left(1-y\right)-2\mathcal{K}\left[x_{B}\left[6-y
\left(6-y\right)\right]-2\left[5-y\left(5-y\right)
\right]\right]\right]\right]\nonumber \\
&  & \times\left[\left[\left[F_{1u}\left(t\right)-F_{1d}
\left(t\right)\right]+\left[F_{2u}\left(t\right)-F_{2d}
\left(t\right)\right]\right]\left[\Re\left(
\mathcal{H}_{WC}^{-}\right)+\Re\left(\mathcal{E}_{WC}^{-}
\right)\right]+g_{A}\left(t\right)\left[\Re\left(
\widetilde{\mathcal{H}}_{WC}^{+}\right)+\Re\left(
\widetilde{\mathcal{E}}_{WC}^{+}
\right)\right]\right.\nonumber \\
&  & \left.+g_{P}\left(t\right)\left[\Re\left(
\widetilde{\mathcal{H}}_{WC}^{+}\right)+\frac{t}{4M^{2}}
\Re\left(\widetilde{\mathcal{E}}_{WC}^{+}
\right)\right]\right]\nonumber \\
&  & +\xi\left[2\mathcal{K}\left(1-y\right)^{2}
\right]\left[\left[\left[F_{1u}\left(t\right)-F_{1d}
\left(t\right)\right]+\left[F_{2u}\left(t
\right)-F_{2d}\left(t\right)\right]\right]\Re\left(
\widetilde{\mathcal{H}}_{WC}^{+}\right)+g_{A}\left(t
\right)\left[\Re\left(\mathcal{H}_{WC}^{-}\right)+
\Re\left(\mathcal{E}_{WC}^{-}\right)
\right]\right]\nonumber \\
&  & +2\left(Q_{u}-Q_{d}\right)\left[
\frac{\left(2-x_{B}\right)\left(1-y\right)}{2x_{B}y}
\left[-\left(2-x_{B}\right)\left(1-y\right)+4
\mathcal{K}\left[1-x_{B}\left(2-y\right)
\right]\right]\right.\nonumber \\
&  & \times\left[\left[F_{1u}\left(t\right)-F_{1d}
\left(t\right)\right]^{2}-\frac{t}{4M^{2}}
\left[F_{2u}\left(t\right)-F_{2d}\left(t\right)
\right]^{2}+g_{A}^{2}\left(t\right)\right]\nonumber \\
&  & +\xi\left[-\frac{\left(1-y\right)}{2y}
\right]\left[-\left(2-x_{B}\right)\left(1-y
\right)+4\mathcal{K}\left(1-x_{B}\right)
\left(2-y\right)\right]\nonumber \\
&  & \left.\left.\times\left[\left[
\left[F_{1u}\left(t\right)-F_{1d}\left(t\right)
\right]+\left[F_{2u}\left(t\right)-F_{2d}
\left(t\right)\right]\right]^{2}+g_{A}\left(t\right)
\left[g_{A}\left(t\right)+g_{P}\left(t\right)
\right]+g_{P}\left(t\right)\left[g_{A}\left(t
\right)+\frac{t}{4M^{2}}g_{P}\left(t\right)
\right]\right]\right]\right\} .\nonumber \\
\end{eqnarray}
\label{eq:interferencewc2}

Due to current conservation, the total amplitude,
\begin{eqnarray}
\mathrm{T}_{\nu n} & = & \mathrm{T}_{C\nu n}+
\mathrm{T}_{BH\nu n}\ ,
\label{eq:totalamplitudewcdvcs}
\end{eqnarray}
is transverse with respect to the momentum $q_{2}$ of the outgoing
real photon. In other words, by replacing the polarization vector
$\epsilon_{\mu}^{*}\left(q_{2}\right)$ in Eqs.~(\ref{eq:TmatrixCWC})
and (\ref{eq:TmatrixBHcharged}) by $q_{2\mu}$, we find for the
Compton contribution:
\begin{eqnarray}
\mathrm{T}_{C\nu n} & \longrightarrow & -\frac{\sqrt{2}}{2}
\left(Q_{u}-Q_{d}\right)\left|e\right|G_{F}\bar{u}
\left(k'\right)\gamma_{\nu}\left(1-\gamma_{5}\right)u
\left(k\right)\bar{u}\left(p_{2},s_{2}\right)
\bigg[\left[F_{1u}\left(t\right)-F_{1d}\left(t\right)\right]
\gamma^{\nu}\nonumber \\
&  & \left.+\left[F_{2u}\left(t\right)-F_{2d}\left(t\right)
\right]\frac{\left(\gamma^{\nu}\not\! r-\not\! r
\gamma^{\nu}\right)}{4M}-g_{A}\left(t\right)
\gamma^{\nu}\gamma_{5}+g_{P}\left(t\right)
\frac{r^{\nu}\gamma_{5}}{2M}\right]u\left(p_{1},s_{1}
\right) ,
\label{eq:Comptonemgaugeinvariance}
\end{eqnarray}
while for the Bethe-Heitler part we have:
\begin{eqnarray}
\mathrm{T}_{BH\nu n} & \longrightarrow & \frac{\sqrt{2}}{2}
\left|e\right|G_{F}\bar{u}\left(k'\right)\gamma^{\nu}\left(1-
\gamma_{5}\right)u\left(k\right)\bar{u}\left(p_{2},s_{2}\right)
\left[\left[F_{1u}\left(t\right)-F_{1d}
\left(t\right)\right]\gamma_{\nu}-\left[F_{2u}\left(t\right)-F_{2d}
\left(t\right)\right]
\frac{i\sigma_{\nu\lambda}r^{\lambda}}{2M}\right.\nonumber \\
&  & \left.-g_{A}\left(t\right)\gamma_{\nu}\gamma_{5}+g_{P}
\left(t\right)\frac{r_{\nu}\gamma_{5}}{2M}\right]u
\left(p_{1},s_{1}\right)\ .
\label{eq:BHemgaugeinvariance}
\end{eqnarray}
Clearly, summing the expressions (\ref{eq:Comptonemgaugeinvariance})
and (\ref{eq:BHemgaugeinvariance}) gives zero, so that the
electromagnetic gauge invariance is explicitly satisfied.

\subsection{Electromagnetic scattering}

For completeness (and     for comparison   with   existing  
results \cite{Belitsky:2001ns}),   we  also consider the standard electromagnetic
DVCS process on a proton target.
The T-matrix for the pure Compton process is given by:
\begin{eqnarray}
\mathrm{T}_{C} & = & \frac{\left|e\right|^{3}}{q_{1}^{2}}
\bar{u}\left(k'\right)
\gamma_{\nu}u\left(k\right)\epsilon_{\mu}^{*}
\left(q_{2}\right)\mathcal{T}_{EM}^{\mu\nu}\ ,
\label{eq:Tmatrixcompton2}
\end{eqnarray}
where $\mathcal{T}_{EM}^{\mu\nu}$ is the reduced electromagnetic
virtual Compton scattering amplitude, which can be easily reproduced
from Eqs.~(\ref{eq:compactweakneutralamplitude}) and
(\ref{eq:integralsofGPDsweakneutralcurrent}) by discarding all
terms accompanied with the ``minus'' OFPDs, and replacing the
weak vector charge by the electric charge, $c_{V}^{f} \to Q_{f}$.
One should also note the additional factor of 2 in the denominator
of the amplitude due to the structure of the vertex $qqZ^{0}$.
The amplitude can then be written as:
\begin{eqnarray}
\mathcal{T}_{EM}^{\mu\nu} & = &
-\frac{1}{2\left(p\cdot q\right)}\left\{ \left[
\frac{1}{\left(p\cdot q_{2}\right)}
\left(p^{\mu}q_{2}^{\nu}+p^{\nu}q_{2}^{\mu}\right)-g^{\mu\nu}
\right]\right.\nonumber \\
&  & \times\left[\mathcal{H}_{EM}^{+}\left(\xi,t\right)\bar{u}
\left(p_{2},s_{2}\right)
\not\! q_{2}u\left(p_{1},s_{1}\right)+
\mathcal{E}_{EM}^{+}\left(\xi,t\right)\bar{u}\left(p_{2},s_{2}\right)
\frac{\left(\not\! q_{2}
\not\! r-\not\! r\not\! q_{2}\right)}{4M}u\left(p_{1},s_{1}\right)
\right]\nonumber \\
&  & +\left[\frac{1}{\left(p\cdot q_{2}\right)}i
\epsilon^{\mu\nu\rho\eta}q_{2\rho}p_{\eta}
\right]\nonumber \\
&  & \left.\times\left[\widetilde{\mathcal{H}}_{EM}^{+}\left(\xi,t\right)
\bar{u}\left(p_{2},s_{2}\right)\not\! q_{2}
\gamma_{5}u\left(p_{1},s_{1}\right)-\widetilde{\mathcal{E}}_{EM}^{+}
\left(\xi,t\right)
\frac{\left(q_{2}\cdot r\right)}{2M}
\bar{u}\left(p_{2},s_{2}\right)\gamma_{5}u\left(p_{1},s_{1}\right)
\right]\right\} ,
\label{eq:compactEMamplitude}
\end{eqnarray}
where the corresponding integrals, known as Compton form factors,
are given by:
\begin{eqnarray}
\mathcal{H}_{EM}^{+}\left(\xi,t\right) & \equiv &
\sum_{f}Q_{f}^{2}\int_{-1}^{1}\frac{dx}{
\left(x-\xi+i0\right)}H_{f}^{+}\left(x,\xi,t\right)=
\sum_{f}Q_{f}^{2}\int_{-1}^{1}dx\; H_{f}\left(x,\xi,t
\right)\left(\frac{1}{x-\xi+i0}+
\frac{1}{x+\xi-i0}\right)\;,\\
\mathcal{E}_{EM}^{+}\left(\xi,t\right) &
\equiv & \sum_{f}Q_{f}^{2}\int_{-1}^{1}
\frac{dx}{\left(x-\xi+i0\right)}E_{f}^{+}
\left(x,\xi,t\right)=\sum_{f}Q_{f}^{2}
\int_{-1}^{1}dx\; E_{f}\left(x,\xi,t\right)
\left(\frac{1}{x-\xi+i0}+\frac{1}{x+\xi-i0}\right)\;,\\
\widetilde{\mathcal{H}}_{EM}^{+}\left(\xi,t
\right) & \equiv & \sum_{f}Q_{f}^{2}\int_{-1}^{1}
\frac{dx}{\left(x-\xi+i0\right)}\widetilde{H}_{f}^{+}
\left(x,\xi,t\right)=\sum_{f}Q_{f}^{2}\int_{-1}^{1}dx\;
\widetilde{H}_{f}\left(x,\xi,t\right)
\left(\frac{1}{x-\xi+i0}-\frac{1}{x+\xi-i0}\right)\;,\\
\widetilde{\mathcal{E}}_{EM}^{+}\left(\xi,t
\right) & \equiv & \sum_{f}Q_{f}^{2}
\int_{-1}^{1}\frac{dx}{\left(x-\xi+i0\right)}
\widetilde{E}_{f}^{+}\left(x,\xi,t\right)=
\sum_{f}Q_{f}^{2}\int_{-1}^{1}dx\;\widetilde{E}_{f}
\left(x,\xi,t\right)\left(\frac{1}{x-\xi+i0}-
\frac{1}{x+\xi-i0}\right)\;.
\label{eq:integralsofGPDsDVCS}
\end{eqnarray}
Here the spin-averaged square of the Compton T-matrix, $\mathrm{T}_{C}$,
can be written as:
\begin{eqnarray}
\overline{\left|\mathrm{T}_{C}\right|^{2}} & = &
\frac{\left(4\pi\alpha\right)^{3}}{Q_{1}^{4}}L_{
\nu\beta}^{C}H_{C}^{\nu\beta}\ ,
\label{eq:comptonamplitudesquaredDVCS}
\end{eqnarray}
where the electron tensor, after neglecting the electron mass,
becomes:
\begin{eqnarray}
L_{\nu\beta}^{C} & = & 2
\left[k_{\nu}k'_{\beta}+k_{\beta}k'_{\nu}-g_{\nu\beta}
\left(k\cdot k'\right)\right]\ ,
\label{eq:leptonictensorDVCS}
\end{eqnarray}
and the electromagnetic hadronic tensor is:
\begin{eqnarray}
H_{C}^{\nu\beta} & = & -\frac{1}{2}\mathcal{T}_{EM}^{\mu\nu}
\left(\mathcal{T}_{\mu EM}^{\beta}\right)^{*} \nonumber \\
& = & -\mathcal{C}_{1EM}\left[g^{\nu\beta}-
\frac{1}{\left(p\cdot q_{2}\right)}
\left(p^{\nu}q_{2}^{\beta}+p^{\beta}q_{2}^{\nu}
\right)+\frac{M^{2}}{\left(p\cdot q_{2}\right)^{2}}
\left(1-\frac{t}{4M^{2}}\right)q_{2}^{\nu}q_{2}^{\beta}
\right]\ ,
\label{eq:hadronictensorDVCS}
\end{eqnarray}
with
\begin{eqnarray}
\mathcal{C}_{1EM} & = & \left(1-\xi^{2}\right)
\left(\left|\mathcal{H}_{EM}^{+}\right|^{2}+
\left|\widetilde{\mathcal{H}}_{EM}^{+}\right|^{2}
\right)-\left(\xi^{2}+\frac{t}{4M^{2}}\right)
\left|\mathcal{E}_{EM}^{+}\right|^{2}-\xi^{2}
\frac{t}{4M^{2}}\left|\widetilde{\mathcal{E}}_{EM}^{+}
\right|^{2}-2\xi^{2}\Re\left(\mathcal{H}_{EM}^{+*}
\mathcal{E}_{EM}^{+}+\widetilde{\mathcal{H}}_{EM}^{+*}
\widetilde{\mathcal{E}}_{EM}^{+}\right). \nonumber \\
\label{eq:coefficientsemhadronictensor}
\end{eqnarray}
The contraction of tensors then yields:
\begin{eqnarray}
\overline{\left|\mathrm{T}_{C}\right|^{2}} & = &
\frac{\left(4\pi\alpha\right)^{3}}{Q_{1}^{2}y^{2}}2
\left[1+\left(1-y\right)^{2}\right]\mathcal{C}_{1EM}.
\label{eq:contractionemCompton}
\end{eqnarray}

As in electron scattering via the $Z^{0}$-exchange, the
Bethe-Heitler contribution emerges from both diagrams (b) and (c)
in Fig.~\ref{weakprocess}:
\begin{eqnarray}
\mathrm{T}_{BH} & = & \frac{\left|e\right|^{3}}{t}
\epsilon_{\mu}^{*}\left(q_{2}\right)\bar{u}
\left(k'\right)
\left[\frac{\gamma^{\mu}\not\! q_{2}\gamma^{\nu}+2k'^{\mu}
\gamma^{\nu}}{2\left(k'\cdot q_{2}
\right)}+
\frac{-\gamma^{\nu}\not\! q_{2}\gamma^{\mu}+2
\gamma^{\nu}k^{\mu}}{-2\left(k\cdot q_{2}\right)}
\right]u\left(k\right)\left\langle p\left(p_{2},s_{2}\right)
\right|J_{\nu}^{EM}\left(0\right)
\left|p\left(p_{1},s_{1}
\right)\right\rangle ,
\label{eq:TmatrixBH2}
\end{eqnarray}
where the proton matrix element of the electromagnetic transition
current is parametrized in terms of the usual Dirac and Pauli
proton form factors:
\begin{eqnarray}
\left\langle p\left(p_{2},s_{2}\right)\right|J_{\nu}^{EM}
\left(0\right)\left|p\left(p_{1},s_{1}\right)
\right\rangle  & = & \bar{u}\left(p_{2},s_{2}\right)
\left[F_{1p}\left(t\right)\gamma_{\nu}-F_{2p}
\left(t\right)\frac{i\sigma_{\nu\lambda}r^{\lambda}}{2M}
\right]u\left(p_{1},s_{1}\right)\ .
\label{eq:protonmatrixelementemdvcs}
\end{eqnarray}
Consequently one can write the Bethe-Heitler squared T-matrix as:
\begin{eqnarray}
\overline{\left|\mathrm{T}_{BH}\right|^{2}} & = &
\frac{\left(4\pi\alpha\right)^{3}}{t^{2}}L_{BH}^{
\nu\beta}H_{\nu\beta}^{BH}\ ,
\label{eq:BHamplitudesquaredDVCS1}
\end{eqnarray}
with the electron tensor:
\begin{eqnarray}
L_{BH}^{\nu\beta} & = & -\frac{1}{2}\mathrm{Tr}\left\{ \not\! k'
\left[\frac{\gamma^{\mu}\not\! q_{2}\gamma^{\nu}+2k'^{\mu}
\gamma^{\nu}}{2\left(k'\cdot q_{2}\right)}+
\frac{\gamma^{\nu}\not\! q_{2}\gamma^{\mu}-2k^{\mu}
\gamma^{\nu}}{2\left(k\cdot q_{2}\right)}\right]\not\! k
\left[\frac{\gamma^{\beta}\not\! q_{2}\gamma_{\mu}+2
\gamma^{\beta}k'_{\mu}}{2\left(k'\cdot q_{2}\right)}+
\frac{\gamma_{\mu}\not\! q_{2}\gamma^{\beta}-2
\gamma^{\beta}k_{\mu}}{2\left(k\cdot q_{2}\right)}
\right]\right\} \nonumber \\
& = & \frac{2}{\left(k'\cdot q_{2}\right)}
\left[k^{\nu}q_{2}^{\beta}+k^{\beta}q_{2}^{\nu}-g^{\nu\beta}
\left(k\cdot q_{2}\right)\right]+
\frac{2}{\left(k\cdot q_{2}\right)}
\left[k'^{\nu}q_{2}^{\beta}+k'^{\beta}q_{2}^{\nu}-g^{\nu\beta}
\left(k'\cdot q_{2}\right)\right]\nonumber \\
&  & +\frac{2}{\left(k'\cdot q_{2}\right)
\left(k\cdot q_{2}\right)}\left[\left(k\cdot q_{2}
\right)\left[k^{\nu}k'^{\beta}+k^{\beta}k'^{\nu}+2k'^{
\nu}k'^{\beta}\right]-\left(k'\cdot q_{2}\right)
\left[k^{\nu}k'^{\beta}+k^{\beta}k'^{\nu}+2k^{\nu}k^{\beta}
\right]\right.\nonumber \\
&  & \hspace*{0.5cm}\left.+\left(k\cdot k'\right)
\left[k^{\nu}q_{2}^{\beta}+k^{\beta}q_{2}^{\nu}-k'^{
\nu}q_{2}^{\beta}-k'^{\beta}q_{2}^{\nu}+2k^{\nu}k'^{
\beta}+2k^{\beta}k'^{\nu}\right]+2g^{\nu\beta}
\left(k\cdot k'\right)\left[\left(k'\cdot q_{2}\right)-
\left(k\cdot q_{2}\right)-\left(k\cdot k'
\right)\right]\right]\ , \nonumber \\
\label{eq:BHleptonictensorDVCS}
\end{eqnarray}
and the hadronic tensor:
\begin{eqnarray}
H_{\nu\beta}^{BH} & = & \frac{1}{2}\sum_{s_{1},s_{2}}
\left\langle p\left(p_{2},s_{2}\right)
\right|J_{\nu}^{EM}\left(0\right)
\left|p\left(p_{1},s_{1}\right)\right\rangle \left\langle p
\left(p_{2},s_{2}\right)\right|J_{\beta}^{EM}
\left(0\right)\left|p\left(p_{1},s_{1}\right)\right
\rangle ^{*}\nonumber \\
& = & t\left[g_{\nu\beta}-\frac{r_{\nu}r_{\beta}}{t}
\right]\left[F_{1p}\left(t\right)+F_{2p}
\left(t\right)\right]^{2}+4\left[p_{1\nu}-\frac{r_{\nu}}{2}
\right]\left[p_{1\beta}-\frac{r_{\beta}}{2}\right]
\left[F_{1p}^{2}\left(t\right)-
\frac{t}{4M^{2}}F_{2p}^{2}\left(t\right)\right]\ .
\label{eq:BHhadronictensorDVCS}
\end{eqnarray}
The squared Bethe-Heitler amplitude is then found to be:
\begin{eqnarray}
\overline{\left|\mathrm{T}_{BH}\right|^{2}} & = & -
\frac{\left(4\pi\alpha\right)^{3}}{t}\left.
\left[\frac{4}{2\kappa\left(2\kappa+\tau+1
\right)x_{B}^{2}y^{2}}\right]\right\{ x_{B}^{2}y^{2}
\left[\left(2\kappa+1\right)^{2}+\left(2\kappa+
\tau\right)^{2}\right]\left[F_{1p}\left(t
\right)+F_{2p}\left(t\right)\right]^{2}\nonumber \\
&  & +\frac{2}{\tau}\left[2\tau\left(1-y
\right)-x_{B}y\tau\left(4\kappa+\tau+1
\right)+y^{2}\left[\tau+x_{B}\tau\left(2\kappa+
\tau\right)+x_{B}^{2}\mu\left[1+8\kappa^{2}+
\tau^{2}+4\kappa\left(1+\tau\right)
\right]\right]\right]\nonumber \\
&  & \left.\times\left[F_{1p}\left(t\right)^{2}-
\frac{t}{4M^{2}}F_{2p}
\left(t\right)^{2}\right]\right\}\ .
\label{eq:BHamplitudesquaredDVCS2}
\end{eqnarray}

Finally, for the Compton--Bethe-Heitler interference part, we find:
\begin{eqnarray}
\mathcal{I} & = & \frac{\left(4\pi\alpha
\right)^{3}}{4Q_{1}^{2}t}\left\{ \mathrm{Tr}
\left\{ \not\! k'\gamma_{\nu}\not\! k\left[
\frac{\gamma^{\beta}\left(\not\! q_{2}
\gamma_{\mu}+2k'_{\mu}\right)}{2\left(k'\cdot q_{2}\right)}+
\frac{\left(\gamma_{\mu}\not\! q_{2}-2k_{\mu}\right)
\gamma^{\beta}}{2\left(k\cdot q_{2}\right)}\right]
\right\} \right.\nonumber \\
&  & \times\mathrm{Tr}\left\{ \left(\not\! p_{2}+M
\right)\hat{\mathcal{T}}_{EM}^{\mu\nu}\left(
\not\! p_{1}+M\right)\left[F_{1p}\left(t\right)
\gamma_{\beta}+F_{2p}\left(t\right)\frac{i
\sigma_{\beta\tau}r^{\tau}}{2M}\right]
\right\} \nonumber \\
&  & +\mathrm{Tr}\left\{ \not\! k\gamma_{\beta}
\not\! k'\left[\frac{\left(\gamma^{\mu}
\not\! q_{2}+2k'^{\mu}\right)\gamma^{\nu}}{2
\left(k'\cdot q_{2}\right)}+\frac{\gamma^{\nu}
\left(\not\! q_{2}\gamma^{\mu}-2k^{\mu}
\right)}{2\left(k\cdot q_{2}\right)}
\right]\right\} \nonumber \\
&  & \left.\times\mathrm{Tr}\left\{ \left(
\not\! p_{1}+M\right)\left(\hat{
\mathcal{T}}_{\mu EM}^{\beta}\right)^{*}
\left(\not\! p_{2}+M\right)\left[F_{1p}
\left(t\right)\gamma_{\nu}-F_{2p}\left(t\right)
\frac{i\sigma_{\nu\lambda}r^{\lambda}}{2M}
\right]\right\} \right\} \nonumber \\
& = & \frac{\left(4\pi\alpha\right)^{3}}{t}
\left[\frac{1}{2\kappa\left(2\kappa+\tau+1
\right)x_{B}y^{3}}\right]\left[8\mathcal{K}
\left(2-2y+y^{2}\right)\right]\nonumber \\
&  & \times\left\{ F_{1p}\left(t\right)
\Re\left(\mathcal{H}_{EM}^{+}\right)-
\frac{t}{4M^{2}}F_{2p}\left(t\right)
\Re\left(\mathcal{E}_{EM}^{+}\right)+\xi
\left[F_{1p}\left(t\right)+F_{2p}\left(t\right)
\right]\Re\left(\widetilde{\mathcal{H}}_{EM}^{+}
\right)\right\} .
\label{eq:interferenceem}
\end{eqnarray}
It is worth noting that the analytic expressions for the Compton and 
interference contributions, see Eqs.~(\ref{eq:contractionemCompton}) 
and (\ref{eq:interferenceem}), explicitly agree with the results of 
Ref.~\cite{Belitsky:2001ns}. As for the Bethe-Heitler contribution,
one can recover the Bethe-Heitler result of Ref.~\cite{Belitsky:2001ns} 
by substituting Eq.~(\ref{eq:kappa}) into the expression 
(\ref{eq:BHamplitudesquaredDVCS2}).

Having presented the formulas for cross sections for the
electromagnetic and weak neutral and charged current DVCS,
in the next section we compute these cross sections numerically,
using the model for OFPDs in Sec.~\ref{sec:model}.

\subsection{Cross section results}

\begin{figure}[H]
\begin{center}
\includegraphics[%
  scale=0.75]{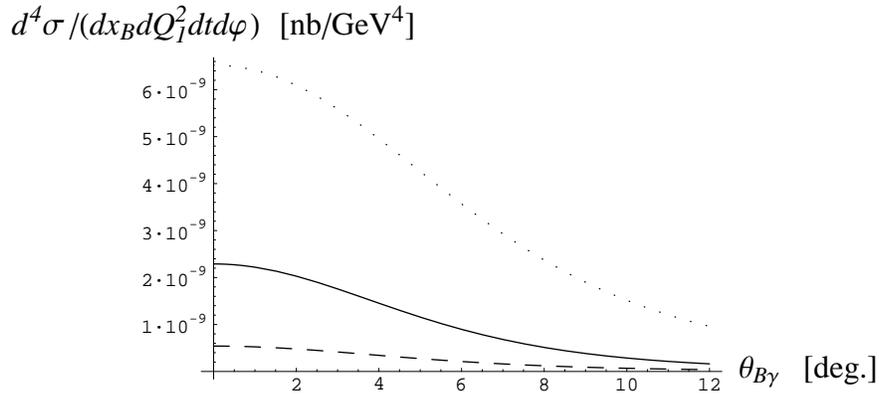}
\end{center}\caption{Compton scattering cross sections for the weak DVCS
	processes corresponding to neutrino-proton scattering via
	the weak neutral current (solid), electron-proton scattering
	via the weak neutral current (dashed), and neutrino-neutron
	scattering via the weak charged current (dotted).
	The cross sections are plotted as a function of the scattering
	angle $\theta_{B\gamma}$ between the incoming virtual weak
	boson and outgoing real photon in the target rest frame for
	$Q_{1}^{2}=2.5\;\mathrm{GeV}^{2}$ and $x_{B}=0.35$ with an
	$\omega=20\;\mathrm{GeV}$ lepton beam. }
\label{weakdvcsCompton}
\end{figure}

The angular dependence of the Compton contribution to the unpolarized
differential cross section for the weak neutral and charged DVCS
processes is illustrated in Fig.~\ref {weakdvcsCompton}.
The results are presented for the region $\theta_{B\gamma}\leq12^\circ$,
which corresponds to taking the invariant momentum transfer
$-t < -1\;\mathrm{GeV}^{2}$ (recall that the requirement for DVCS is that
$-t$ should be much smaller than $Q_{1}^{2}=2.5\;\mathrm{GeV}^{2}$).
The cross sections are observed to be of the same order in magnitude,
and fall off smoothly with increasing $\theta_{B\gamma}$.
For neutrino scattering, the $\nu n$ charged current cross section
is larger than the $\nu p$ neutral current cross section.
Furthermore, in the weak neutral current sector, the cross section
is considerably larger for neutrinos rather than electrons.
The latter reflects the difference in the structure of the leptonic
tensors $L_{\nu\beta}^{\left(\nu\right)}$ and
$L_{\nu\beta}^{\left(e\right)}$ in Eqs.~(\ref{eq:neutrinotensorweakDVCS})
and (\ref{eq:leptonictensor1}), respectively, since the weak neutral
hadronic tensor $H_{WN}^{\nu\beta}$ in Eq.~(\ref{eq:hadronictensorWN})
is the same in both cases.

\begin{figure}[H]
\begin{center}
\includegraphics[%
  scale=0.75]{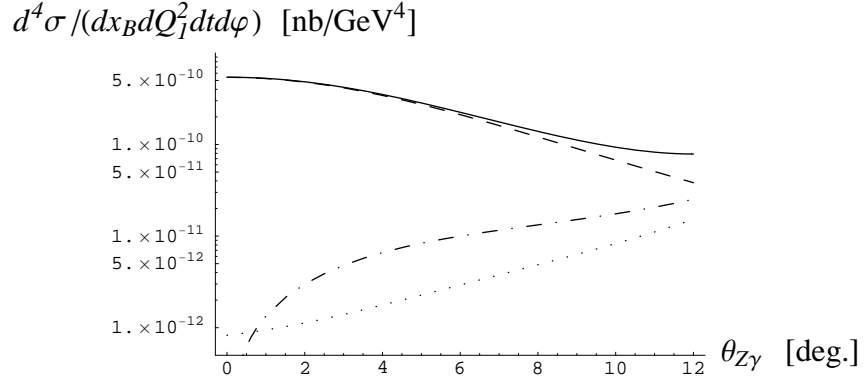}
\end{center}\caption{Compton (dashed), Bethe-Heitler (dotted),
        magnitude of interference (dash-dotted) and total
        (solid) cross sections for electron-proton scattering via the
	weak neutral current plotted as a function of the scattering
	angle between the incoming virtual weak boson and outgoing real
	photon in the target rest frame for
	$Q_{1}^{2}=2.5\;\mathrm{GeV}^{2}$ and $x_{B}=0.35$, with an
	$\omega=20\;\mathrm{GeV}$ electron beam.}
\label{weakneutraldvcsComptonandBH}
\end{figure}

\begin{figure}[H]
\begin{center}
\includegraphics[%
  scale=0.75]{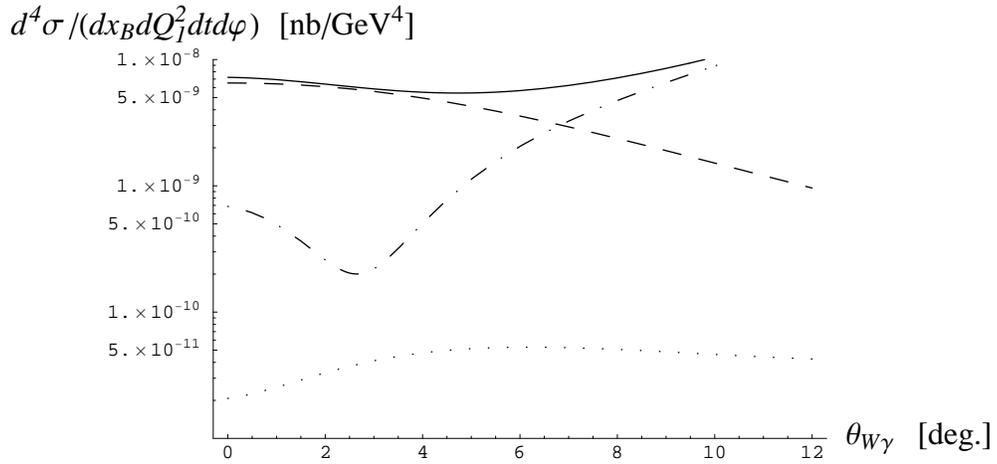}
\end{center}\caption{As in Fig.~\ref{weakneutraldvcsComptonandBH},
	but for the neutrino-neutron charged current cross section.}
\label{weakchargeddvcsComptonandBH}
\end{figure}

\begin{figure}[H]
\begin{center}
\includegraphics[%
  scale=0.75]{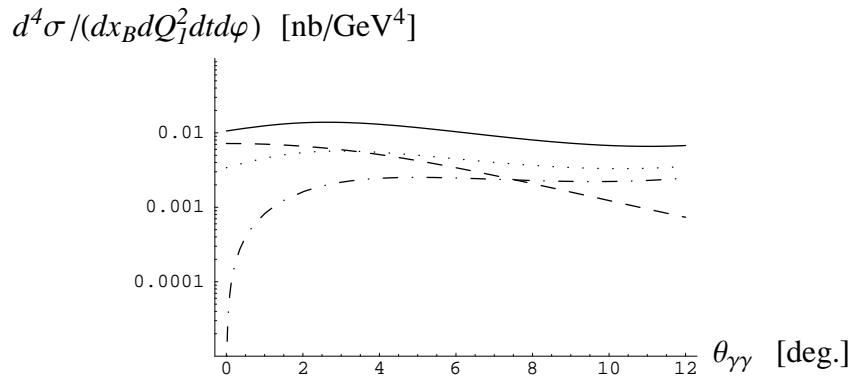}
\end{center}\caption{As in Fig.~\ref{weakchargeddvcsComptonandBH},
	but for the electromagnetic electron-proton cross section.}
\label{emdvcsComptonandBH}
\end{figure}

To investigate the magnitude of the Compton contributions relative
to the corresponding Bethe-Heitler backgrounds, we plot both cross
sections together on a logarithmic scale in
Figs.~\ref{weakneutraldvcsComptonandBH} and
\ref {weakchargeddvcsComptonandBH}.
Note that for the Bethe-Heitler cross section, one should expect
poles when the scattering angle $\theta_{B\gamma}$ coincides with
$\phi=20.2^\circ$ and $\phi'=25.3^\circ$, which corresponds to the
outgoing real photon being collinear with either the incoming or
scattered lepton.
In both cases, the Compton cross section is significantly larger
than the Bethe-Heitler background, especially at small angles.
This is to be contrasted with the electromagnetic current case,
Fig.~\ref{emdvcsComptonandBH}, in which the Bethe-Heitler contribution
is larger than the Compton for $\theta_{\gamma\gamma} > 3^\circ$.

\begin{figure}[H]
\begin{center}
\includegraphics[%
  scale=0.75]{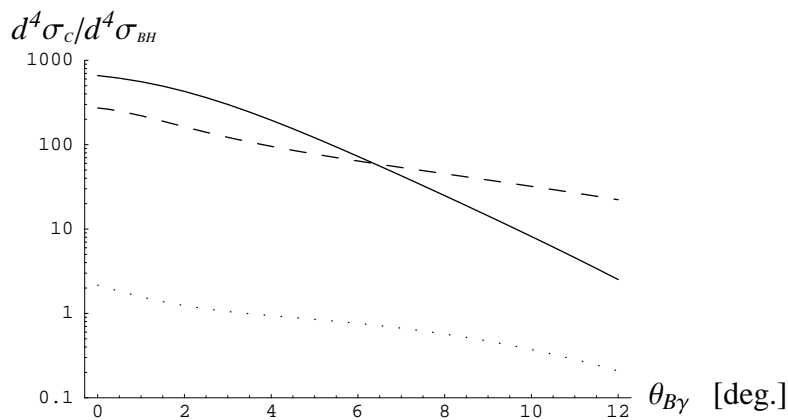}
\end{center}\caption{Ratio of Compton and Bethe-Heitler cross sections
	for weak neutral (solid), weak charged (dashed) and
	electromagnetic (dotted) DVCS, as a function of the scattering
	angle between the incoming boson $B=Z^{0}$, $W^{+}$ or
	$\gamma$ and the outgoing real photon in the target rest frame,
	for $Q_{1}^{2}=2.5\;\mathrm{GeV}^{2}$, $x_{B}=0.35$ and
	$\omega=20\;\mathrm{GeV}$.}
\label{ratioComptonBH}
\end{figure}

The relative importance of the various contributions is even more
graphically illustrated in Fig.~\ref{ratioComptonBH}, where we plot
the ratio of the Compton and Bethe-Heitler cross sections for the
weak neutral current (solid), weak charged current (dashed), and
electromagnetic (dotted) DVCS.
In the forward direction (e.g. for $\theta_{B\gamma}\leq4^\circ$ with
the ratio $-t/Q_{1}^{2}\leq0.1$), the Bethe-Heitler contribution is
strongly suppressed by a factor more than $100$ compared to the Compton
contribution for both weak neutral and weak charged current
scattering. This is in contrast with the electromagnetic current case,
where the ratio between the cross sections remains of order unity.
Although based on a rather simple model of OFPDs, these results
should be helpful in providing some guidance for future neutrino
scattering experiments.
%

\section{Conclusions\label{conclusions}}

In this work we have presented a comprehensive analysis of DVCS-like
processes induced by the exchange of weak vector bosons.
Weak DVCS is an important new tool for studying the quark structure
of the nucleon, providing complementary information to that available
through electromagnetic probes.

As in elastic form factor studies, or in deep inelastic scattering,
weak DVCS provides access to different combinations of GPDs to those
which can appear in ordinary DVCS.
The flavor dependence of the weak charges, for example, means that
one has more sensitivity to $d$ quarks in the proton than in
electromagnetic scattering.
Furthermore, the $V-A$ nature of the weak interaction allows both
$C$-odd and $C$-even combinations of GPDs to be measured, thereby
providing a direct separation of the valence and sea content of GPDs.
Since DVCS is sensitive to spin-averaged as well as spin-dependent
GPDs, this feature means that one can extract spin-dependent valence
and sea quark distributions without polarizing the target.
The only other means of obtaining spin-dependent $C$-odd distributions
would be through neutrino-polarized nucleon scattering, which is
prohibitive, however, because of the large quantities of nuclear
material which would need to polarized.
An additional feature of weak DVCS is that it enables one to access
GPDs that are nondiagonal in quark flavor, such as those associated
with the neutron-to-proton transition.

In the present study, we have derived the weak virtual Compton
scattering amplitude for the neutral current, in both neutrino and
charged lepton scattering, as well as for the charged current.
The amplitudes have been calculated in the leading, twist-2
approximation using the light-cone expansion of the current product,
in terms of QCD string operators in coordinate space.

To quantify our results, and to explore the feasibility of measuring
weak DVCS cross sections, we have used a simple, factorized model to
estimate the cross sections in kinematics relevant to future
high-intensity neutrino experiments.
In contrast to the standard electromagnetic DVCS process, we find that
at small scattering angles the Compton signal is enhanced relative to
the corresponding Bethe-Heitler contribution.
This should make contamination from the Bethe-Heitler backgrounds
less of a problem when extracting the weak DVCS signal.

While the current model analysis has been exploratory, in future one
can use more elaborate models for nucleon GPDs, including sea quark
effects, and contributions from the plus and minus distributions
separately.
In the small-$|t|$ region of DVCS kinematics, it would also be of
interest to examine further the contribution from the pion pole,
through the $\widetilde{E}_{f}$ distribution.
Finally, it will be necessary to extend the approach to include
twist-3 terms in order to apply the formalism at moderate energies,
where the suppression of higher twist contributions is not guaranteed.
The results presented here provide an important starting point in
realizing the program of extracting GPDs in neutrino scattering.

\begin{acknowledgments}

We   are  grateful  to R. Ransome  for   attracting   our 
attention   to neutrino DVCS,
and  to D. M{\"u}ller  for an   instructive 
discussion  of  the  twist     decomposition
for   interference   terms. 
Work of  A.P.  and   A.R.  was    supported in part by 
DOE   Grant DE-FG02-97ER41028. 

{\it Notice:} Authored by Jefferson Science Associates, 
LLC under U.S. DOE Contract
No. DE-AC05-06OR23177. The U.S. Government retains a non-exclusive,
paid-up, irrevocable, world-wide license to publish or reproduce this
manuscript for U.S. Government purposes.

\end{acknowledgments}


\end{document}